\documentclass[fleqn,usenatbib]{mnras}

\usepackage[T1]{fontenc}

\usepackage{ae,aecompl}

\DeclareRobustCommand{\VAN}[3]{#2}
\let\VANthebibliography\thebibliography
\def\thebibliography{\DeclareRobustCommand{\VAN}[3]{##3}\VANthebibliography}

\usepackage{graphicx}	
\usepackage{amsmath}	
\usepackage{amssymb}	

\usepackage{ulem}
\usepackage{soul}
\usepackage{natbib}

\usepackage{xcolor}

\newcommand{\msun}{$\rm M_\odot$}

\title[UDGs in different environments]{Origin and evolution of ultra-diffuse galaxies in different environments}

\author[J. A. Benavides et al.]{
Jos\'e A. Benavides$^{1,2}$\href{https://orcid.org/0000-0003-1896-0424}{\includegraphics[scale=0.8]{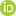}}\thanks{E-mail: jose.benavides@unc.edu.ar},
Laura V. Sales$^{3}$\href{https://orcid.org/0000-0002-3790-720X}{\includegraphics[scale=0.8]{figures/orcid.png}}, 
Mario. G. Abadi$^{1,2}$\href{https://orcid.org/0000-0003-3055-6678}{\includegraphics[scale=0.8]{figures/orcid.png}},
Federico Marinacci$^{4}$\href{https://orcid.org/0000-0003-3816-7028}{\includegraphics[scale=0.8]{figures/orcid.png}}, \newauthor
Mark Vogelsberger$^{5}$\href{https://orcid.org/0000-0001-8593-7692}{\includegraphics[scale=0.8]{figures/orcid.png}} and Lars Hernquist$^{6}$
\\
$^{1}$Instituto de Astronom\'ia Te\'orica y Experimental, CONICET-UNC, Laprida 854, X5000BGR, C\'ordoba, Argentina\\
$^{2}$Observatorio Astron\'omico de C\'ordoba, Universidad Nacional de C\'odoba, Laprida 854, X5000BGR, C\'ordoba, Argentina\\
$^{3}$Department of Physics and Astronomy, University of California, Riverside, CA, 92521, USA\\
$^{4}$Department of Physics and Astronomy "Augusto Righi", University of Bologna, I-40129 Bologna, Italy\\ 
$^{5}$Department of Physics, Massachusetts Institute of Technology, Cambridge, MA 02139, USA \\
$^{6}$Institute for Theory and Computation, Harvard-Smithsonian Center for Astrophysics, Cambridge, MA 02138, USA\\ 
}

\date{Accepted XXX. Received YYY; in original form ZZZ}

\pubyear{2022}

\begin{document}
\label{firstpage}
\pagerange{\pageref{firstpage}--\pageref{lastpage}}
\maketitle

\begin{abstract}
We study the formation of ultra-diffuse galaxies (UDGs) using the cosmological hydrodynamical simulation TNG50 of the Illustris-TNG suite. We define UDGs as dwarf galaxies in the stellar mass range $\rm{7.5 \leq log (M_{\star} / M_{\odot}) \leq 9 }$ that are in the $5\%$ most extended tail of the simulated mass-size relation. This results in a sample of UDGs with half-mass radii $\rm{r_{h \star } \gtrsim 2 \ kpc}$ and surface brightness between $\rm{24.5}$ and $\rm{28 \ mag \ arcsec^{-2}}$, similar to definitions of UDGs in observations. The large cosmological volume in TNG50 allows for a comparison of UDGs properties in different environments, from the field to galaxy clusters with virial mass $\rm{M_{200} \sim 2 \times 10^{14} ~ M_{\odot}}$. All UDGs in our sample have dwarf-mass haloes ($\rm{M_{200}\sim 10^{11} ~ M_{\odot} }$) and show the same environmental trends as normal dwarfs: field UDGs are star-forming and blue while satellite UDGs are typically quiescent and red. The TNG50 simulation predicts UDGs that populate preferentially higher spin haloes and more massive haloes at fixed $\rm{M_{\star}}$ compared to non-UDG dwarfs. This applies also to most satellite UDGs, which are actually ``born" UDGs in the field and infall into groups and clusters without significant changes to their size. We find, however, a small subset of satellite UDGs ($\lesssim 10 \%$) with present-day stellar size a factor $\geq 1.5$ larger than at infall, confirming that tidal effects, particularly in the lower mass dwarfs, are also a viable formation mechanism for some of these dwarfs, although subdominant in this simulation.  

\end{abstract}




\section{Introduction}

Ultra diffuse galaxies (UDGs) are an extreme case of low-surface brightness galaxies \citep{Sandage1984, Impey1988, Dalcanton1997, deBlok1997, McConnachie2008, Conselice2018} with luminosities in the regime of dwarfs $L=[\sim 10^{7}$-$10^{9}]\; \rm L_\odot$ and extended half-light radii $\rm{R_{e} \geq 1.5 ~ kpc}$. UDGs were first detected in large numbers in the Coma cluster \citep{vanDokkum2015a}, followed by several pioneering works mining for these dim dwarfs in galaxy groups and clusters \citep[e.g., ][]{Koda2015, vanDokkum2015a, vanDokkum2015b,  Mihos2015, vanderBurg2016, Yagi2016, ManceraPina2019a, Lim2020,Venhola2022,LaMarca2022}. 

These studies confirmed that up to thousands of UDGs can be found in single individual clusters and that the abundance of UDGs scales close to linearly with host halo mass \citep{vanderBurg2016,ManceraPina2018,Lee2020}. UDGs are therefore a numerous population in high-density environments and contribute significantly to well-studied statistics such as the satellite luminosity function or galaxy clustering. They are also a fundamental component of our understanding of dwarf galaxy formation, as some of their observed properties, such as size, globular cluster (GC) content or inferred dynamical mass, remain difficult to reconcile with theoretical models \citep{Sales2022}.

One of the most striking aspects of early UDG studies was the discovery of a large number of associated GCs \citep{vanDokkum2016,Peng2016,vanDokkum2017}, which together with their survival in high density environments motivated the idea that they live in overly massive dark matter haloes, being more comparable to Milky-Way like objects than to dwarfs. Follow up studies have shown that this is not necessarily the case for all UDGs, some of which could include a few or even no GCs at all \citep{Beasley2016,Amorisco2018,Lim2018,Saifollahi2021,Saifollahi2022}, as well as a wide range of dynamical mass estimates \citep{Beasley2016,Toloba2018, vanDokkum2019, Danieli2019,Doppel2021,Gannon2022}. 

The combination of current results on GC content and kinematic mass estimates suggests that a large fraction of the UDGs inhabit dwarf-mass haloes like regular dwarfs, instead of being comparable to $L_{\star}$ objects as originally thought \citep[see brief review in ][]{Trujillo2021}. Currently, the dark matter content of UDGs remains an interesting topic of debate, with some cases of extreme mass-to-light ratios (Toloba et al., {\it in-prep}) or the overabundance of GCs in some UDGs being particularly intriguing aspects of their formation (see \citealt{Trujillo-Gomez2022}, \citealt{Carleton2021} and \citealt{Danieli2022} for possible mechanisms to explain high GC numbers).

Several theoretical models were crafted to explain the formation of UDGs with large stellar sizes, which can be roughly divided into three main categories: internal processes, externally-driven processes or a combination of both. Internal processes include high angular momentum \citep{Amorisco2016,Rong2017} or bursty and prolonged star formation with their associated breathing-mode stellar outflows \citep{DiCintio2017, Chan2018} as main drivers of the extended sizes in UDGs. Environmentally-driven mechanisms include the expansion of otherwise normal dwarf galaxies due to different processes such as tidal heating \citep{Jiang2019}, tidal stripping \citep{Carleton2019, Maccio2021}, non-adiabatic expansion of the stars due to gas removal \citep{Safarzadeh2017}, mergers \citep{Wright2021} or stellar dimming after star formation truncation induced by cluster environment \citep{Tremmel2020}. 

The third class of UDG-formation models invoke the need for a {\it combination} of internal and external processes. For instance, in simulations presented in \cite{Jiang2019} and \citet{Sales2020}, the UDG population in group and cluster-like objects consists of the infall of extended dwarfs already ``born'' UDGs in the field plus the addition of newly formed UDGs due to tidal stripping of more massive galaxies \citep{Sales2020} or tidal heating of normal dwarfs \citep{Jiang2019}, with the inclusion of both mechanisms necessary to reproduce the observed number of UDGs in groups and clusters. Another example of a mixed origin is presented in \citet{Jackson2021}, where low-surface brightness galaxies (some of which would qualify as UDGs) are formed by a combination of early assembly due to high-density regions followed by stripping and tidal perturbations driven by the environment at late times.    

While all UDG formation mechanisms discussed above may play a role in shaping these galaxies to some degree, the identification of a main driver for UDG formation is still elusive and, most importantly, the predictions from these theoretical models can be mutually contradictory. For instance, early analytical models predict that UDGs form in haloes with high-spin \citep{Amorisco2016,Rong2017}, however several simulations teams find no particular bias in the halo spin of UDGs vs. no-UDGs dwarfs \citep{Jiang2019,Tremmel2020,Wright2021}. The formation time for UDGs is also poorly constrained in theoretical models, with some results from cosmological simulations suggesting early assembly \citep{Jackson2021, Wright2021} but semi-analytical models favouring instead late formation times \citep{Rong2017}, partially confirmed later by \citet{Kong2022} for the case of gas-rich UDGs in the field. 

With the important caveat in mind that different definitions have been applied in the past to identify UDGs in observations and simulations, and that this can have a significant effect on the conclusions drawn \citep{VanNest2022}, the lack of consensus coming from different simulations is most likely also tracking differences in the baryonic treatment and feedback model adopted in each numerical experiment, which has been shown to impact considerably the structural properties of galaxies in simulations \citep{Sales2010,Scannapieco2012}. 

A promising avenue to help break degeneracies between predictions from different models is to compare populations of UDGs formed across different environments. For instance, UDG formation models that are purely environmentally driven would expect a much smaller population of UDGs in the field. Encouragingly, observational efforts have already rendered large samples of field UDGs candidates, defined as galaxies in low density environments that satisfy similar surface densities and radius cuts as the traditional UDGs in clusters \citep[e.g., ][]{MartinezDelgado2016,Roman2017b,Leisman2017,Jones2018,Roman2019,Jones2021,Marleau2021}. 

In the overwhelming majority, field UDGs are blue and star forming, in contrast with group and cluster UDGs, which are red and quiescent \citep{Prole2021,Kadowaki2021,Zaritsky2022, ManceraPina2019a, vanderBurg2016}. Note that a handful of identified field-UDGs are also found to be red and quiescent \citep[e.g., ][]{MartinezDelgado2016,Papastergis2017,Roman2019}, which may be naturally explained  through backsplash orbits \citep{Benavides2021}. Is there a link between gas-rich UDGs in the field and their quiescent counterparts found today in group and cluster environments?

Kinematically, the (admittedly scarce) available data from a sample of HI-rich field UDGs suggests very low inner dark matter density \citep{Jones2018, ManceraPina2019b, ManceraPina2022}, which disfavours the existence of very massive dark matter haloes in these dwarfs, in agreement with dwarf-mass halo estimates in several quiescent UDGs. We caution though that these results should be carefully interpreted, as the sample with resolved rotation curves is small and caveats on the inclination determination and possible missalignments might apply \citep{Read2016b,Oman2016,Gault2021,Sellwood2022}. Most importantly, while we know little about the kinematics of the field UDGs, we know even less about their GC content given the difficulties in identifying GC-like objects in lumpy star-forming disks. This lack of common ground to compare observations of UDGs in the field and in clusters makes it very challenging to trace a possible evolutionary link between star-forming UDGs and quiescent UDGs in groups and clusters using observational samples.\\

Numerical simulations, with their ability to trace objects across time, are an ideal tool to tackle such questions and provide important guidance to future observations. However, because of numerical resolution demands, UDG studies in cosmological simulations have been limited in the past mostly to zoom-in field dwarfs {\it or} zoom-ins of groups and clusters \citep[e.g., ][]{DiCintio2017,Chan2018,CardonaBarrero2020}, but not to both environments simultaneously. A few efforts have combined environments by studying different simulations using the same feedback prescription (e.g., \citealt{Wright2021} and \citealt{Tremmel2020}, or \citealt{Jiang2019}), but often at different numerical resolutions or still limiting the numbers of objects in one or the other environment.\\

Our work builds on those lines by using the high-resolution cosmological TNG50 numerical simulation \citep{Pillepich2019,Nelson2019TNG}, which uniformly samples a $\sim 50$ Mpc side box to study the formation of UDG galaxies. We present one of the first studies that include UDGs in a wide range of environments, from field dwarfs to clusters with virial mass $\sim 10^{14}~$\msun\; representing fairly intermediate structures such as filaments and backsplash regions. The relatively large volume simulated in TNG50 also allows for a uniform sampling of halo formation histories for dwarf UDGs in the field, without biases introduced in selecting individual haloes in zoom-in runs. Most importantly, our sample contains galaxies formed under a unified baryonic treatment, equal numerical resolution independent of the environment and a unified selection criteria, simplifying the interpretation of our results and possible comparison to current and future observations.\\

The paper is organised as follows: In Sec.~\ref{sec:sims} we briefly describe the simulation and discuss the selection of UDGs applied in our sample. In Sec.~\ref{sec:form} we analyze the halo and stellar properties of UDGs comparing different environments. We study in detail the evolution of satellite UDGs in Sec.~\ref{sec:sats} and determine the role of tidal effects on defining their extended sizes. We discuss and summarise our results in \ref{sec:concl}. 

\section{Simulations and method}
\label{sec:sims}

We use the TNG50 cosmological hydrodynamical simulation \citep{Pillepich2019,Nelson2019TNG}, which is the highest resolution box available within the IllustrisTNG project \citep{Pillepich2018a, Pillepich2018b,Nelson2018,Naiman2018,Marinacci2018, Springel2018,Weinberger2018, Nelson2019TNG}. TNG50 follows the evolution of a $\sim 52^3$ Mpc volume set up initially with a total of $ 2 \times 2160^3 $ of gas elements and dark matter particles. The simulation is run using the {\sc arepo} code \citep{Springel2010} to evolve from redshift $z=127$ to the present-day and assumes a set of cosmological parameters consistent with the \citet{PlankColaboration2016} measurements ($ \Omega_m = \Omega_{dm} + \Omega_b = 0.3089 $, $ \Omega_b = 0.0486 $, cosmological constant $ \Omega_{\Lambda} = 0.6911 $ , Hubble constant $\rm{ H_0 = 100 \, h \, km \, s^{-1} \, Mpc^{-1} }$, $ h = 0.6774 $, $ \sigma_8 = 0.8159 $ and spectral index $ n_s = 0.9667 $). The typical mass resolution achieved in TNG50-1, the highest resolution run available for this box and the one analysed here, is $\rm{ m_{bar} } = 8.5 \times 10^4 \, \rm {M_{\odot}} $ and $\rm{ m_{drk} }= 4.5 \times 10^5 \, \rm {M_{\odot}}$, with typical gravitational softening $\epsilon^{z=0}_{DM, \star} = 0.29$ kpc. The average time between snapshot outputs is $\sim 0.14 ~ \rm Gyr$.

The baryonic treatment included in TNG50 is largely based on the previous Illustris project \citep{Vogelsberger2013,Vogelsberger2014a} with some modifications to the stellar and AGN feedback prescriptions described mostly in \cite{Pillepich2018a} and \cite{Weinberger2017}, respectively. Briefly, gas is allowed to cool up to a temperature $T = 10^4\; \rm K$ following the cooling and heating rates computed from local density, redshift and metallicity. Gas above a density $n=0.13$ $\rm cm^{-3}$ is modelled via an equation of state to describe a dual-phase gas \citep{SpringelHernquist2003}. 

Star formation may occur for gas cells above a given density threshold for star formation set at $n_H \simeq 0.1$ $\rm cm^{-3}$. Stellar particles are born assuming a Chabrier initial mass function \citep{Chabrier2003} and their posterior stellar evolution following the prescription described in \citep{Pillepich2018a}. The simulation includes a modelling for momentum input due to stellar feedback as well as metal deposition from evolved stars into the inter-stellar medium. Black hole feedback is also implemented via energy input for both high and low accretion rates, although its modelling is not thought to play a major role in the dwarf mass range analysed in our sample.

The identification of the haloes and subhaloes is done through Friends-of-Friends \citep[FoF]{Davis1985} and {\sc subfind} \citep{Springel2001, Dolag2009}. The evolution of objects through time is followed using the SubLink merger trees \citep{RodriguezGomez2015}. Virial mass, radius and circular velocity ($\rm{M_{200}}$, $\rm{r_{200}}$ and $\rm{V_{200}}$ respectively) are measured using the radius within which the average density is $200$ times the critical density of the universe. The TNG50 box includes a wide range of environments, with the most massive halo having $\rm{M_{200}} \approx 2 \times 10^{14}~$\msun\; followed by a few dozen group-like environment haloes with $ 12.5 < \log (\rm{M_{200}} /\rm  M_{\odot}) < 14 $) and thousands of galaxy- and dwarf-mass objects in the field. 

We will use the term {\it central} to refer to galaxies that are sitting at the center of the potential wells of a given FoF group, and {\it satellite} to refer to anything associated with a FoF group that is not a central. Broadly, we will assume that central galaxies reside in the field, while satellites might belong to a galaxy-, group- or cluster- environment according to the virial mass of their host FoF halo. Infall times for satellites, $t_{\rm inf}$ are defined as the last snapshot where the progenitor is identified as a central galaxy. Properties related to galaxies, such as stellar or gas mass, angular momentum, colors and star formation rates are computed using all particles within the ``galaxy radius", defined here to be twice the half-mass radius of the stars $\rm{r_{h,\star}}$, a common assumption in previous works from the Illustris and Illustris-TNG projects.

\begin{figure}
\centering
\includegraphics[width=\columnwidth]{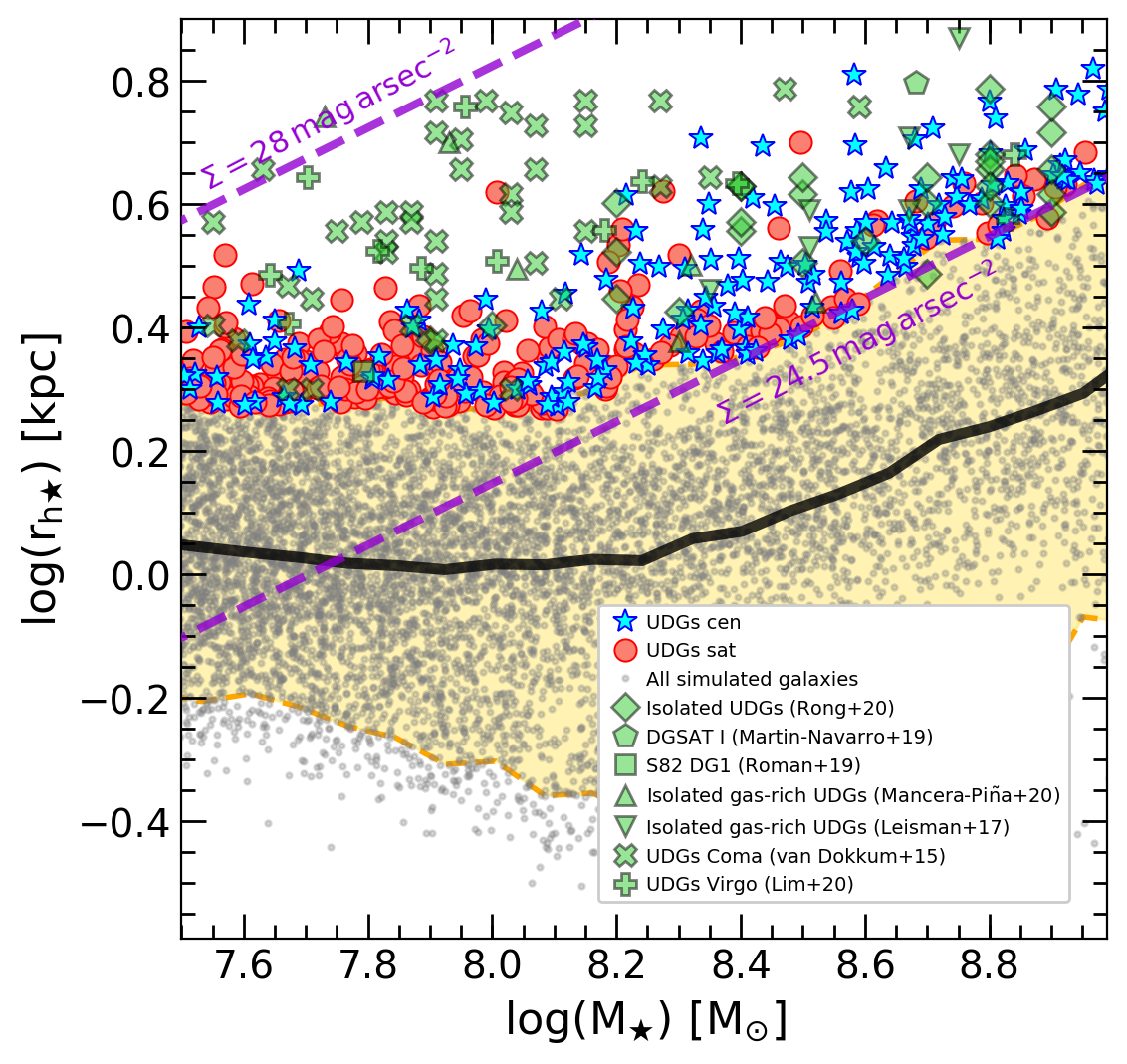}
\caption{Relation between stellar mass and stellar size (defined as the half mass radius, $\rm{r_{h \star}}$) for all simulated galaxies in  the mass range $\rm{ \log (M_{\star} / M_{\odot}) =  [7.5, 9.0] } $ in the TNG50 simulation (grey dots). The median relation is highlighted with the thick solid black line. UDGs are defined as the most extended $5\%$ of the sample at each stellar mass, and highlighted in red circles or blue stars for satellites or centrals, respectively. For comparison, we define the population of normal dwarfs to be all simulated galaxies with sizes between the $5^{\rm th}$ and $95^{\rm th}$ percentile (yellow region). Thin violet dashed lines indicate lines of constant surface brightness assuming a mass-to-light ratio equal to unity and encompass all our UDG sample. Several observational data are shown in black edged symbols, where we transform 2D sizes $R_{\rm eff}$ to 3D assuming $\rm{r_{h \star} = 4/3 \, R_{\rm eff}}$ \citep{Hernquist1990}. Observational data are represented with green smooth symbols: diamonds indicate star-forming UDGs in low-density environments \citep{Rong2020b}; circle is the relatively isolated DGSAT I \citep{MartinNavarro2019}; pentagon is UDG S82-DG-1, an isolated quiescent UDG \citep{Roman2019}, triangles (up and inverts) correspond to gas-rich isolated UDGs \citep{ManceraPina2020, Leisman2017}; crosses are UDGs in the Virgo cluster \citep{Lim2020} and x-symbols for the Coma cluster \citep{vanDokkum2015a}. Our UDG definition agrees well with observational samples.}
\label{fig:mass_size}
\end{figure}

\subsection{Sample of UDGs}
\label{ssec:sample}

We focus on the regime of dwarf galaxies with stellar masses $\rm{M_{\star}}=10^{7.5}$-$10^9$\msun, which in TNG50 means that the lowest mass dwarfs in our sample are resolved with $\sim 570$ stellar particles. In addition, we impose a minimum dark matter mass  $\rm{ M_{DM} \geq 5 \times 10^7 \, M_{\odot} }$ and stellar half-mass radius $\rm{ r_{h \star} \geq 0.3 \, kpc}$ (or effective radius) to remove spurious contamination from baryonic clumps and other numerical artifacts. Fig.~\ref{fig:mass_size} shows the stellar mass - size relation for dwarfs in TNG50 that fulfil these selection criteria. Throughout this article, we will use the word ``size" to characterize the stellar effective radius of galaxies. 

In this work, we identify UDGs as dwarf galaxies with extended sizes that are above the $95^{\rm th}$ percentile of the sample at a given stellar mass, and refer as ``normal dwarfs" to all galaxies within $5^{\rm th}$-$95^{\rm th}$ percentiles in the mass-size relation, indicated by the yellow shaded area in Fig.~\ref{fig:mass_size}. UDGs are highlighted with blue stars or red circles according to whether they are field or satellite objects, respectively. To guide the eye, we include two constant surface brightness lines corresponding roughly to $\Sigma=24.5$ and $\Sigma=28$ mag/arcsec$^2$ (measured within the effective radius and assuming mass-to-light ratio of $1$), which describe well typical luminosities of UDGs in the high and low mass end, respectively. 

With our definition, simulated UDGs are in the ballpark of UDGs from observational surveys in different environments, highlighted with green symbols in Fig.~\ref{fig:mass_size}. We show UDGs in Virgo \citep{Lim2020} and Coma \citep{vanDokkum2015a} clusters, along with low-density regions UDGs \citep{Roman2019, Rong2020b, MartinNavarro2019}. While simulations nicely reproduce the range of sizes observed in the high-mass end studied here, low-mass UDGs in simulations are not as extended as some of the UDGs observed in the Coma cluster \citep{vanDokkum2015a}. 

\begin{figure}
\centering
\includegraphics[width=\columnwidth]{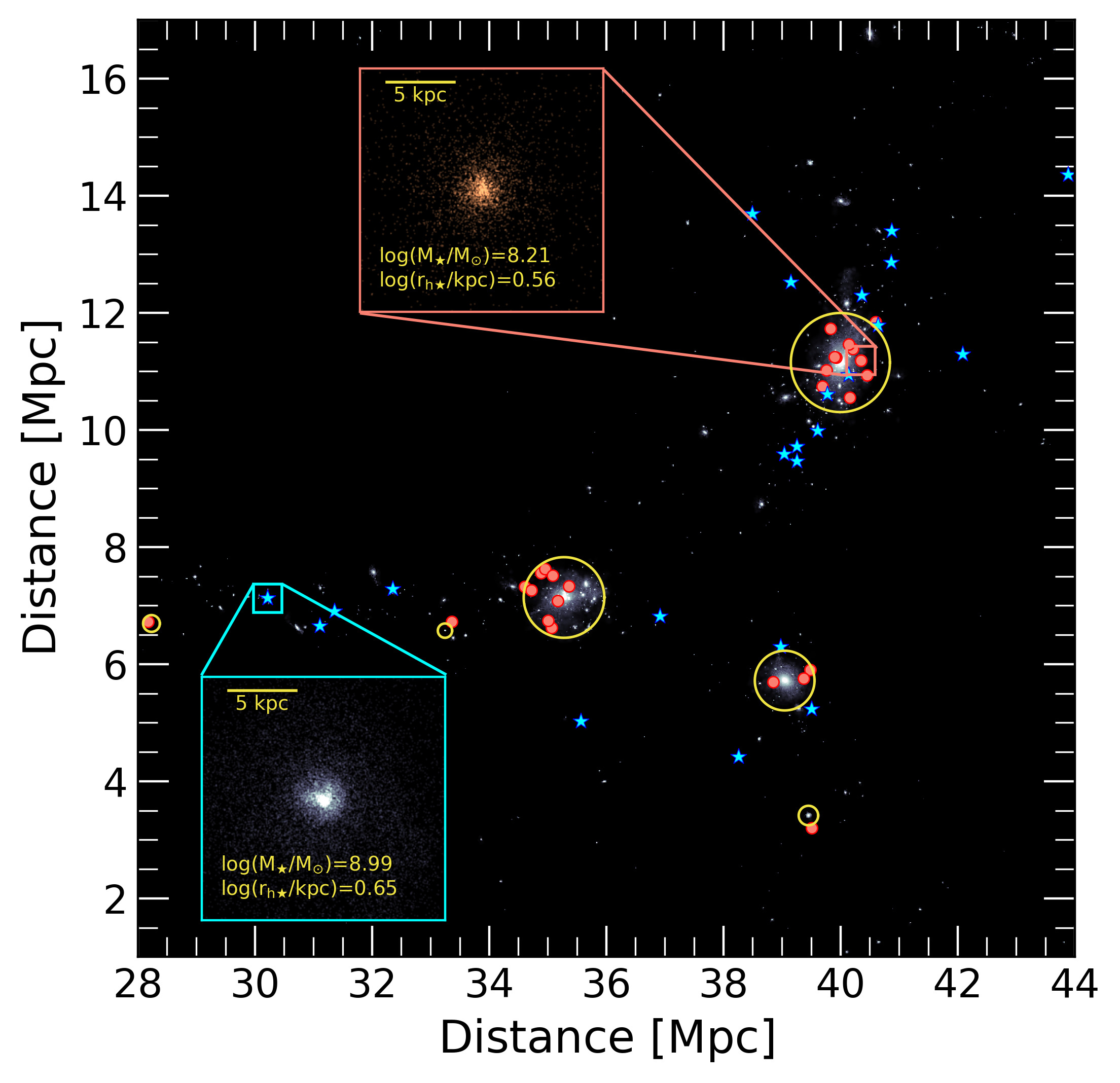}
\caption{Illustration of some of our simulated UDGs and their location within the TNG50 box. Blue stars represent central UDGs (galaxies from the field) and red circles show satellite UDGs (galaxies in groups). The zoom-in panels show the stellar component of two UDG examples, one in the field (bottom left corner) and a satellite of a $\rm{M_{200}} \sim 10^{13}~$\msun\; host (top right). Yellow circles indicate the virial radius of some galaxy- and group-size haloes in this region of the box.}
\label{fig:box}
\end{figure}

Our final sample of simulated UDGs in the $\rm{M_{\star}=10^{7.5}}$-$\rm{10^9 ~ M_{\odot}}$ mass range comprises 176 field objects and 260 satellites. An example of their distribution with respect to other simulated structures in the box is shown in Fig.~\ref{fig:box}. Note that our definition aligns more closely with UDGs being {\it outliers} of the mass-size scaling relation, in a similar fashion as used in the \citet{Lim2020} study of the Virgo cluster, and does not assume a fixed radius or surface brightness cut as preferred in other studies \citep[e.g., ][]{vanDokkum2015a}. Our decision is mostly driven by the mass-dependent behaviour of size in the mass range analysed shown in Fig.~\ref{fig:mass_size}.  As such, our interpretation of UDGs will always be as extreme objects compared to the formation of the majority of dwarfs at the same mass, or ``normal dwarfs'', which represent the $90\%$ of the population. We emphasise that in our definition UDGs can never dominate or become a significant fraction of the dwarf population, but are instead defined as the most extended outliers \citep[see for instance ][ for a different approach]{Tremmel2020,Wright2021,Jackson2021}.

\section{Formation of UDGs in different environments}
\label{sec:form}

We analyse in what follows the main properties of our identified UDGs in relation to non-UDG dwarfs formed in the simulations. 
We focus on quantities that have been proposed in the past as associated to the formation of UDGs: halo mass, spin, mergers and star-formation indicators.  

\subsection{Halo mass}
\label{ssec:halomass}

\begin{figure*}
\centering
\includegraphics[width=1.0\textwidth]{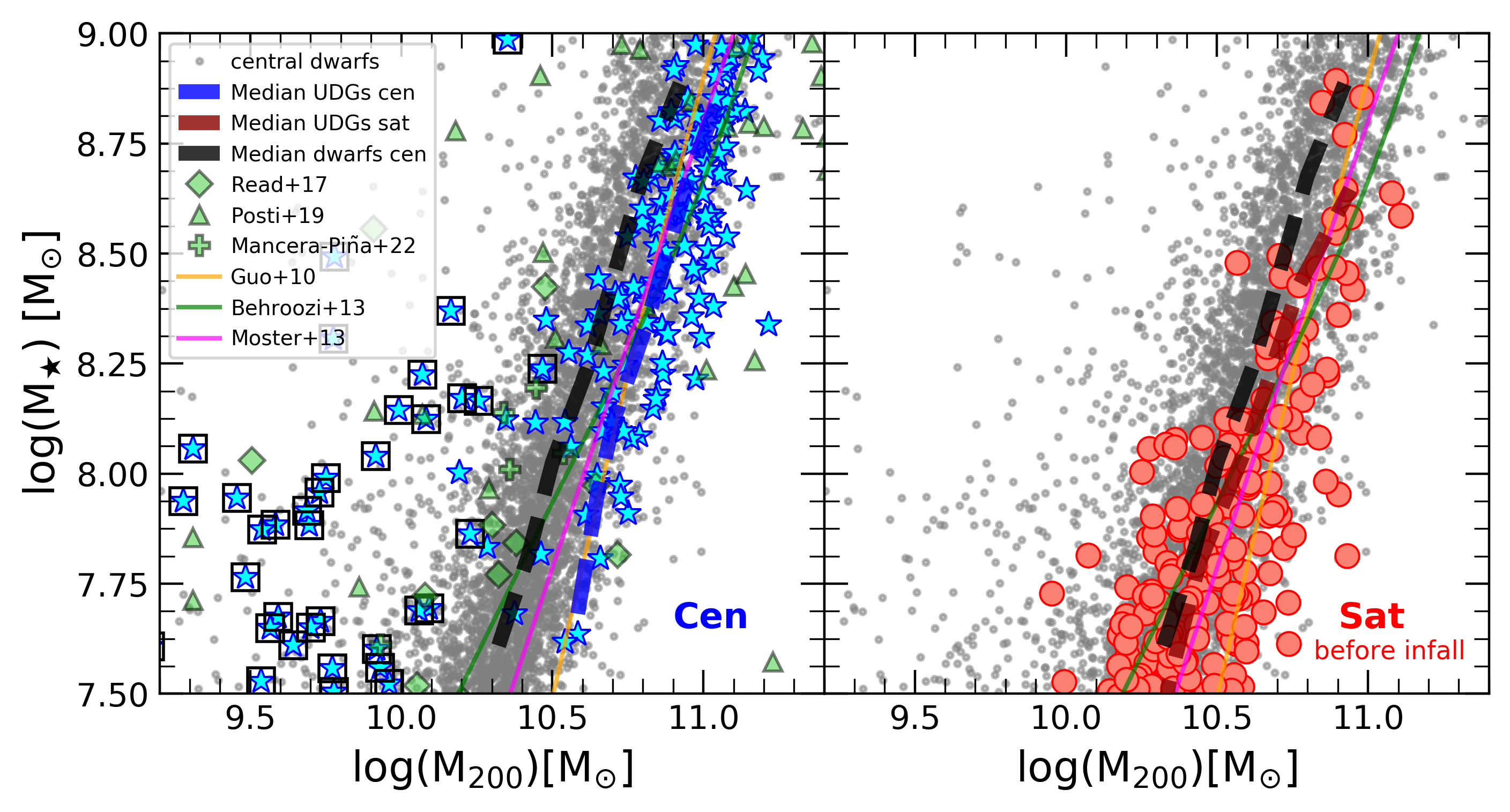}
\caption{Relation between stellar mass and halo mass for the dwarf population in TNG50. As before, blue stars and red circles indicate central (left panel) and satellite (right) UDGs, respectively. For satellite UDGs, we plot their halo mass the last time they were classified as centrals. As reference, grey symbols indicate the population of central normal dwarfs in both panels as well as abundance matching relations from \citet{Guo2010, Behroozi2013, Moster2013}, these being extrapolations below $\rm{M_\star} \sim 10^8~$\msun. Most UDGs follow a similar stellar - halo mass relation as normal dwarfs, with a bias towards larger halo mass at fixed $\rm{M_{\star}}$, which can be seen by the median relations shown in thick long-dashed lines for normal dwarfs (black) and central (blue) or satellite (red) UDGs in left and right panel, respectively.  Note that several outliers appear for this relation, which are related to backsplash objects. We highlight the population of central backsplash quenched UDGs presented in \citet{Benavides2021} with black squares on the left panel. All simulated UDGs have halo masses in the dwarf range, $\rm{M_{200} \leq 10^{11.2} ~ M_{\odot}}$. For reference, we show observational data of dwarf galaxies from \citet{ManceraPina2022b} as green crosses, for a compilation of dwarf galaxies presented in \citet{Read2017} as green diamonds and for the SPARC dwarf sample \citep{Posti2019} with green triangles.}
\label{fig:abundance_matching}
\end{figure*}

Fig.~\ref{fig:abundance_matching} shows the stellar mass - halo mass relation for simulated dwarfs in TNG50. UDGs are highlighted in blue stars or red circles for field (left panel) or satellite (right panel) objects, respectively. Note that in the case of satellites, their virial mass at the present day is ill defined and we therefore use their last recorded virial mass before they joined another FoF group and lost their central status. All central dwarfs (non-UDG) are also included in grey symbols as a comparison. For reference, we include predictions from previous abundance matching models \citet {Guo2010, Behroozi2013, Moster2013}, these being extrapolations below $M_\star \sim 10^8~$\msun, as well as results from observational studies by \citet{Read2017} and \citet{ManceraPina2022b}. 

The first main prediction of our study is that all UDGs in our sample populate dwarf-mass haloes that span the virial mass range $\rm{M_{200}=10^{10}}$-$\rm{10^{11.2} ~ M_{\odot}}$. Note that some seemingly field UDGs in the left panel (highlighted by black squares) may have virial masses below this range and are clear outliers in the stellar mass - halo mass relation. These objects, which were introduced in \citet{Benavides2021} in detail, are backsplash objects that are in the field today but were satellites of more massive systems in the past. As such, their present-day halo mass is significantly reduced from what it was before the interaction as a result of tidal stripping, placing these UDGs outside the main galaxy locus of the simulation. We have checked that the virial mass of these objects before the backsplash interaction was in the virial mass range quoted above for the UDG population.  

Fig.~\ref{fig:abundance_matching} also suggests that at a fixed stellar mass, UDGs populate more massive haloes than non-UDG objects, a trend that seems stronger in more luminous UDGs and in the field, although still true for satellites. The thick dashed black and blue or red lines indicates the median of the normal population, and UDGs, for centrals and satellites, respectively. The scatter upwards in halo mass at fixed stellar mass, in combination with the tight relation between halo mass and globular cluster content \citep{Harris2015,Harris2017}, is interesting and might help explain differences in the globular cluster content of UDGs compared to normal dwarfs of similar luminosity \citep[see e.g., ][]{Trujillo-Gomez2022}. 

Additionally, we have checked that the average dark matter density profiles of field UDGs and non-UDGs are in good agreement with each other, suggesting no significant differences in the concentration parameter of their dark matter halos. This is in principle in contradiction with results in \citet{Kong2022}, which we explain as a result of very different selection criteria: these authors select halos as potential hosts of UDG galaxies (in the dark matter only version of TNG50) by identifying those that reproduce the observed rotation velocity of $6$ field UDGs while we select our sample purely in terms of structural properties of the stellar components. We defer the study on the predicted kinematical properties of our sample to future upcoming work (Doppel et al., {\it in-prep}).

\subsection{Halo Spin}
\label{ssec:spin}

\begin{figure}
\centering
\includegraphics[width=\columnwidth]{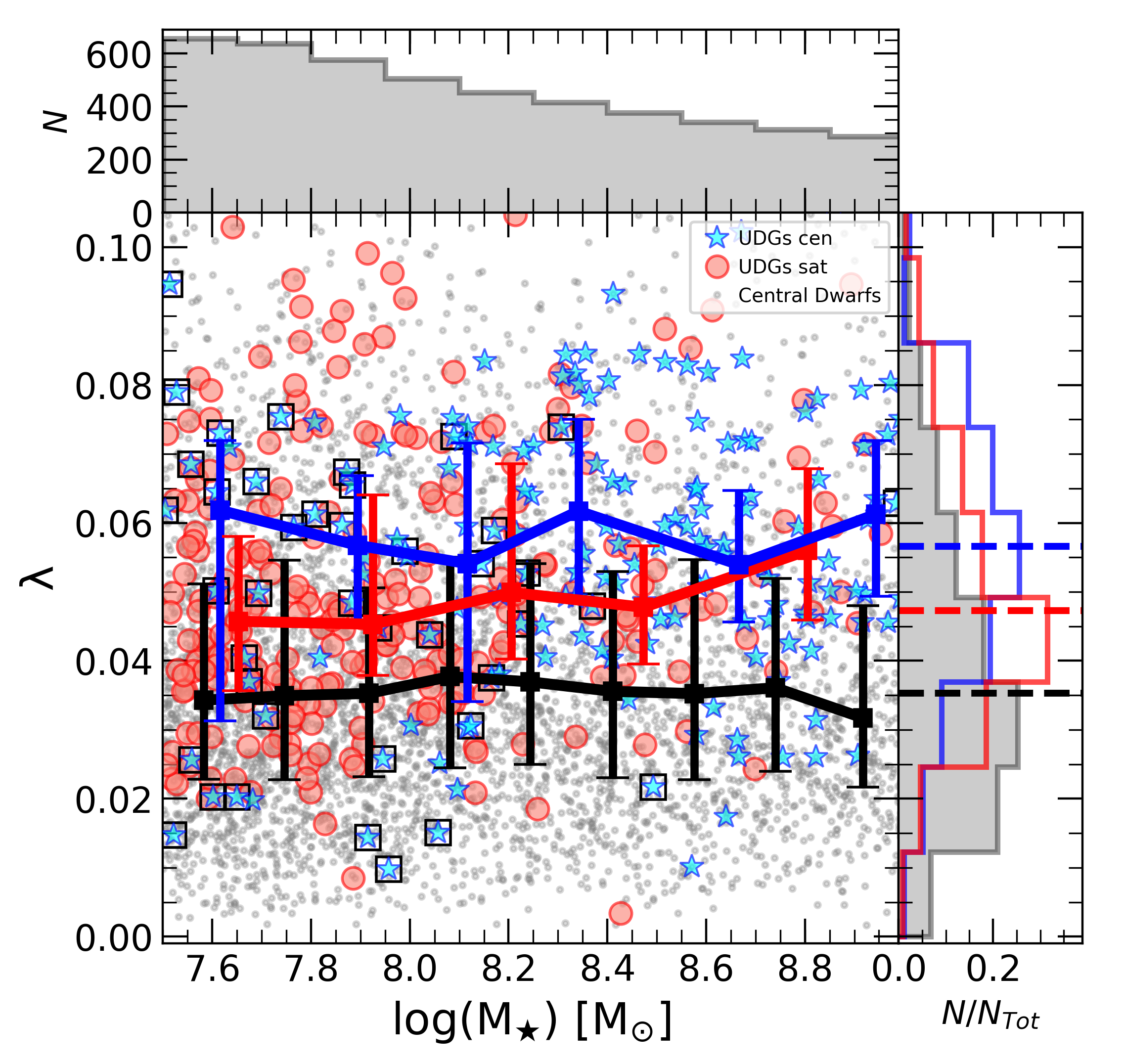}
\caption{Halo spin for dwarf galaxies at a given stellar mass. Normal field dwarfs are shown in grey, while blue stars and red circles highlight central UDGs and satellite UDGs, respectively, with central backsplash UDGs indicated by black squares. For satellite UDGs we measure halo spin at infall since environmental effects may have influenced its present-day value. The median spin at fixed $\rm{M_\star}$ of the normal dwarf population is indicated by the solid black curve and with error bars indicating $25^{\rm th}$-$75^{\rm th}$ percentiles, the average value for all mass bins is $\lambda_{\rm{dwarf}} = 0.035^{+0.017}_{-0.012}$. Thick blue and red lines show the median and $25^{\rm th}$-$75^{\rm th}$  percentiles for the central and satellite UDG population. UDGs occupy preferentially higher-spin haloes. The median and rms dispersion for the UDGs are $\lambda_{\rm{cen}} = 0.059^{+0.012}_{-0.015}$ and $\lambda_{\rm{sat}} = 0.047^{+0.014}_{-0.009}$, for the centrals and satellites population, respectively. The different halo spins in normal vs. UDG population may be better shown in the histograms on the right, with dashed lines indicating the medians of the normal (black), central (blue) and satellites (red) UDGs.}
\label{fig:spin}
\end{figure}

One of the first analytical models for the formation of UDGs postulated that they inhabit dwarf-mass haloes with a high spin parameter \citep{Amorisco2016}. We show in Fig.~\ref{fig:spin} that simulated UDGs in TNG50 indeed are characterised by a higher than average spin $\lambda$, defined as:

\begin{equation}
\lambda =\lambda' = \frac{J}{\sqrt{2} M_{200} V_{200} r_{200} }
\label{eq:lambda}	
\end{equation}

\noindent
where $J$ is the angular momentum within $\rm{r_{200}}$ \citep{Bullock2001}. As before, grey symbols show the distribution of all central dwarfs while field UDGs are highlighted as blue stars (backsplash UDGs marked with black squares). Solid lines show median spin and the 25$^{\rm th}$-75$^{\rm th}$ percentiles of different populations as a function of stellar mass while the vertical side panel shows the $\lambda$ histograms of each sample. 

The median spin of the UDG sample (blue solid line) is $\lambda \approx 0.06$, independent of stellar mass, which is systematically above the median of the whole field population $\lambda \approx 0.035$ (black solid). Notice that the value of the central population as a whole is in agreement with the average spin of dark matter haloes expected in $\Lambda$CDM \citep{Maccio2007}. Satellite UDGs at the present day have their dark matter spin affected by tidal disruption, so we measure their spin at the time of infall and show individual results in red symbols and the median trend with a red solid line. Satellite UDGs also show an excess of angular momentum, with a median $\lambda \approx 0.05$, which is lower than the field UDGs but still biased high with respect to the field dwarf population. These results confirm in simulations some of the previous analytical and semi-analytical models for the formation of UDGs based on high-spin haloes \citep{Amorisco2016,Rong2017}.

\begin{figure}
\centering
\includegraphics[width=\columnwidth]{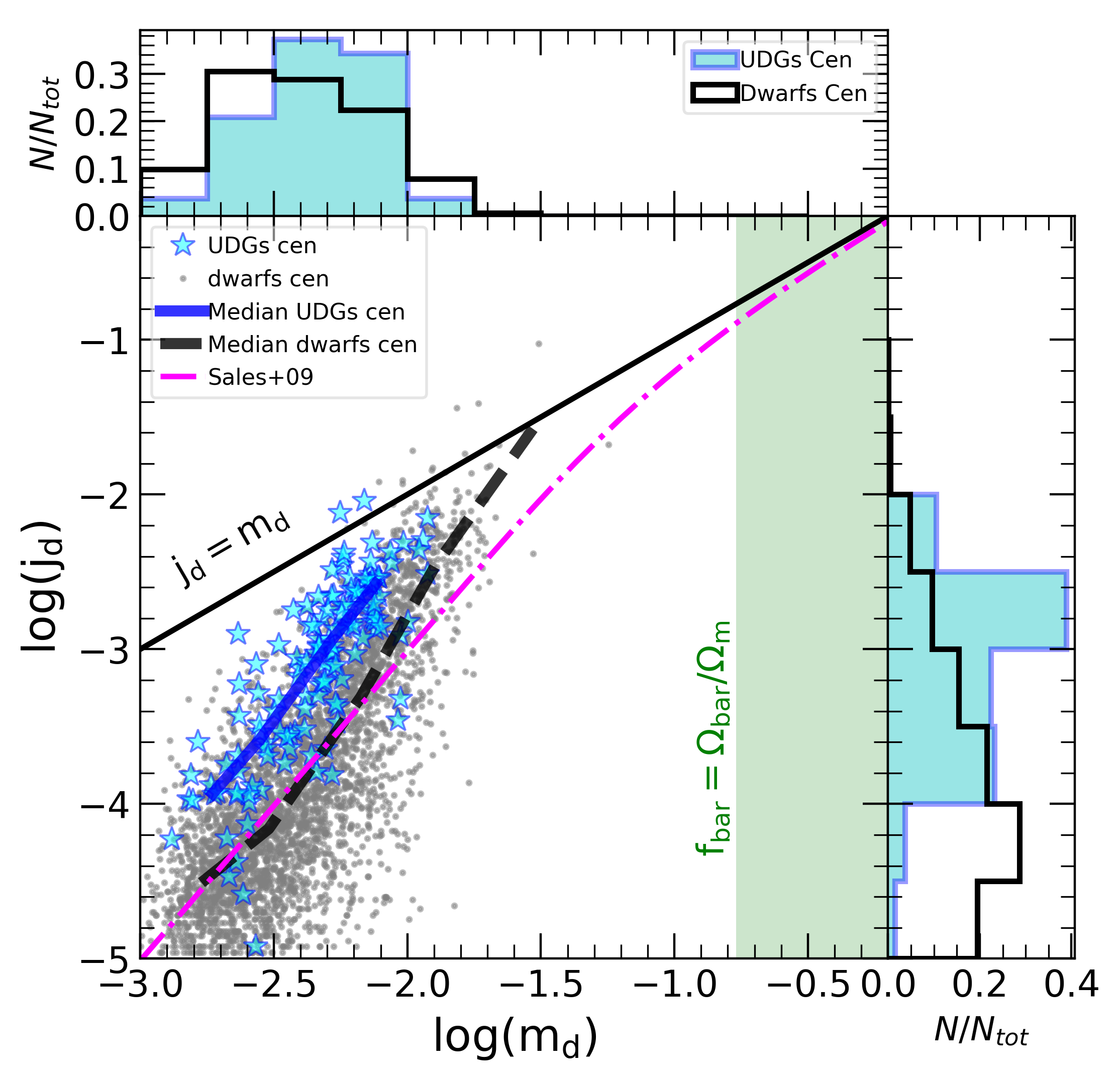}
\caption{Fraction of specific angular momentum retained by the stellar component of the galaxy, $j_d = J_d / J_h$ (with $J_d$ and $J_h$ the angular momentum of the stars in the galaxy and the halo, respectively), as a function of the stellar mass fraction $m_d = M_d/M_h$ (with $M_d$ and $M_h$ the mass in the stellar component of the galaxy and the halo, respectively), following the formalism presented in the \citet{mmw1998} model. Only central galaxies are shown, grey for normal dwarfs and blue stars for simulated central UDGs (without including the population that was stripped and highlighted with black squares in Fig.~\ref{fig:abundance_matching}). Medians for each population are indicated by the dashed black and solid blue lines for non-UDGs and central UDGs, respectively, and show that UDGs retain $\sim 2$ times more specific angular momentum from the halo at a fixed $m_d$ compared to non-UDGs. The black continuous line indicates the $j_d=m_d$ relation, while the magenta line shows the fitting polynomial approximation proposed in \citet{Sales2009} for the OWLS simulations. The vertical green region indicates the limit for the universal baryon fraction $\rm{f_{bar} = \Omega_{bar} / \Omega_m = 0.17}$. The histograms for the central UDGs population are included in both axes.}
\label{fig:mdjd}
\end{figure}

Interestingly, the main panel of Fig.~\ref{fig:spin} shows that not all high-spin halos host UDGs, as suggested by the presence of grey symbols with high $\lambda$ parameters. This means that while the spin is playing a major role, it is not the only defining quantity in forming dwarf galaxies with large radii. Indeed, in idealized analytical models of disk formation\footnote{and assuming disk size (for an exponential profile half-mass radius is equivalent to 1.7 times the scale-length) as a proxy for galaxy size irrespective of morphology.} \citep[]{mmw1998,Somerville2018}, the scale-length of a disk $R_d$ that settles into a surrounding dark matter halo scales linearly with the halo spin parameter but depends on two other fundamental factors: halo virial radius and the ratio $j_d/m_d$:
\begin{equation}
	R_d = \frac{1}{\sqrt 2} \left( \frac{j_d}{m_d} \right) \lambda r_{200}	\ ,
\label{eq:mmw}	
\end{equation}
\noindent
where $j_d=J_d/J_{200}$ is the fraction of the angular momentum in the disk compared to the virial angular momentum and $m_d=M_d/M_{200}$ is the fraction of the mass in the disk compared to that of the halo. This formula assumes an isothermal halo and infinitely thin disk, but additional factors might be added to introduce more complexity, such as a different dark matter profile or the possibility of baryonic contraction \citep{mmw1998}. Eq.~\ref{eq:mmw} provides a useful framework for understanding the results in Fig.~\ref{fig:abundance_matching} and \ref{fig:spin}. The most extended  galaxies (or UDGs) will form preferentially in more massive haloes at a given $\rm{M_{\star}}$ (larger $r_{200}$) and haloes with higher spins $\lambda$. What values of $j_d$ and $m_d$ do simulated UDGs have?

In simpler terms, the ratio $j_d/m_d$ in Eq.~\ref{eq:mmw} measures the fraction of the {\it specific} angular momentum that a galaxy manages to capture from the dark matter halo. While ideally baryons and dark matter may share similar specific angular momentum at the time of decoupling from the general expansion of the Universe, when most of the angular momentum is imprinted \citep{Doroshkevich1970,White1984,Porciani2002a,Porciani2002b}, we know that only a small fraction of the baryons are locked up as stars in galaxies in order to explain results from abundance matching or the zero-point of the Tully-Fisher relation \citep[e.g., ][]{Dutton2012}. How efficient is that small fraction of the baryons to bring most of the angular momentum of the halo (which seems necessary to reproduce the observed galaxy sizes) is controlled by baryonic feedback and galactic outflows \citep[e.g., ][]{Sales2010,Brook2011,Brook2012,Ubler2014}. 

We can use this formalism to gain intuition on galaxy half-mass radius predicted by the simulation, irrespective of the specific morphology and using the full galaxy mass and angular momentum \citep[e.g., see for instance ][]{Sales2009}. We also restrict the analysis to the central population, which is less affected by tidal stripping and the environment. We show in Fig.~\ref{fig:mdjd} the relation between $m_d$ and $j_d$ for the specific feedback and baryonic treatment in TNG50. The ``disk" (e.g. galaxy) mass and angular momentum has been calculated using all stellar particles within twice the half-mass radius of the stars. 

The full dwarf population (grey symbols) is located at quite small values of $m_d \sim 10^{-3}$-$10^{-2}$ (or $\sim 0.5\%$-$5\%$ of all available baryons in the halo), as expected by the inefficiency of star formation in low mass systems. Simulated dwarfs show an increasing fraction of angular momentum $j_d$ in the disk with larger $m_d$ values, which is in a way expected, as incorporating a larger fraction of the baryons presents the opportunity to capture and lock into the galaxy more of the total angular momentum of the halo.

The median $j_d$ at a given $m_d$ computed from all central dwarfs is shown by the dashed black line and follows closely the relation presented in \citet{Sales2009} based on the OWLS simulations \citep{Schaye2010}. The good agreement between these very different sets of simulations is reassuring: while changes in the baryonic treatment used may significantly alter the properties of individual galaxies, the behaviour of different simulations in the $m_d$-$j_d$ plane is more robust to changes in the baryonic physics prescription \citep{Sales2010}. 

Most importantly, Fig.~\ref{fig:mdjd} shows that central UDGs in TNG50 (blue stars) are also outliers in the $m_d$-$j_d$ plane, having captured at a fixed $m_d$ a larger fraction ($\sim 3\times$) of the angular momentum of the halo (median shown as solid blue curve). (Backsplash UDGs have been removed from the sample given their modified halo mass and spins due to previous interactions). The relation between halo spin and $j_d$ is shown in Appendix \ref{app:lambda_and_jd} (Fig.~\ref{fig:jd_vs_lambda}). We conclude that UDGs form as a combination of large halo masses (although still in the dwarf-mass regime), high spins and a higher angular momentum retention in the baryons compared to the halo given their stellar content.

\subsection{Mergers}
\label{ssec:mergers}

\begin{figure}
\centering
\includegraphics[width=\columnwidth]{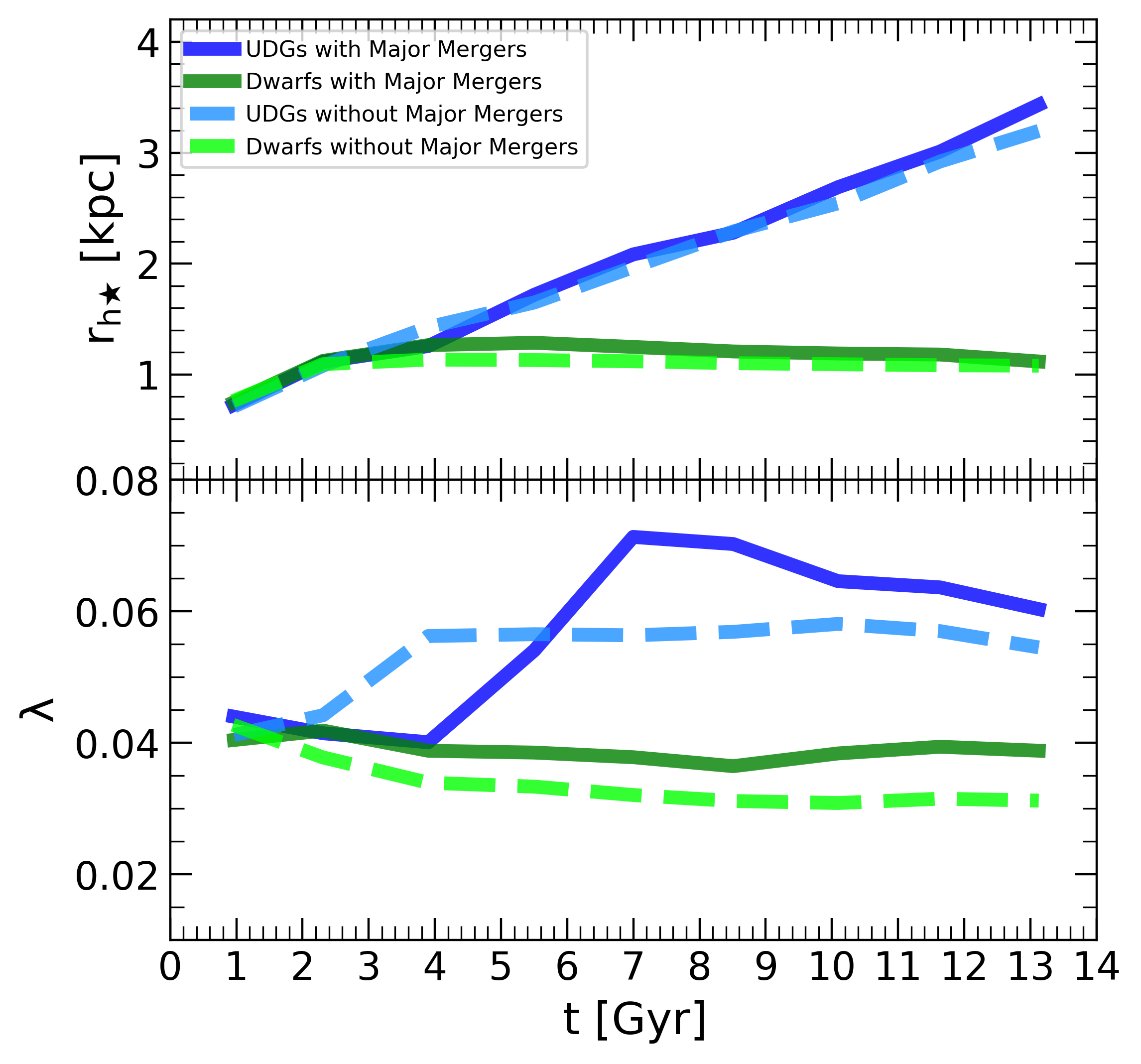}
\caption{Median evolution over time for the sizes (upper panel) and the dimensionless halo spin parameter (lower panel) for field UDGs (blue) and normal dwarf galaxies (green). We divide each sample on those with (dark blue and dark green) and without (light blue and light green) major mergers (with mass ratio $ \mu_{\star} \geq 0.2$). While UDGs are more extended and have higher spin parameters than the non-UDG sample, the presence or not of major mergers does not play a significant role on the median trends, suggesting that mergers are not directly related to the formation of field UDGs in TNG50.}
\label{fig:mergers}
\end{figure}

The connection of mergers to halo spin or angular momentum of the remnant galaxy is complex. But in some cases, when the event is gas-rich and properly aligned, it might help build up galaxies or haloes with high angular momentum content (\citealt{Hopkins2009}, Sotillo et al., in prep.). Merger events could in principle be a channel to deliver high angular momentum gas to the inner regions of haloes to support the formation of extended and low surface brightness galaxies.  Do they play a role in the formation of UDGs in our TNG50 sample? For simplicity of interpretation, we focus the analysis of mergers on the central UDG population since satellites have their late time evolution impacted by their host environment, including the suppression of mergers \citep[see, for instance ][]{Benavides2020}.    

We find no obvious difference in the overall merger history of the central UDGs compared to the non-UDG dwarfs, in agreement with results in \citet{Wright2021}. As shown in Fig.~\ref{fig:histo_lmm} in the Appendix~\ref{app:mergers}, there is a slight tendency for UDGs to have their last major merger (defined as a stellar mass ratio between involved galaxies $\mu_\star \geq 0.2$) at later times than the non-UDG sample. However, the signal is rather weak. Moreover, we also find a fraction of our field UDGs ($\sim 65 \%$) that has never experienced a major merger at all, signalling that mergers are not fundamental to the formation of UDGs in our simulation. 

Fig.~\ref{fig:mergers} shows the median evolution of stellar size (upper panel) and halo spin (bottom) for the UDGs with mergers (blue solid curve) and without major mergers (light blue dashed line). For completeness, we also divide the normal dwarf sample into with and without major mergers (dark solid green and light dashed green curves, respectively). Note that the presence or not of major mergers make no difference in the overall non-UDG or UDG populations. We find that UDGs both with and without mergers have an excess of spin compared to the non-UDG sample, reinforcing the link between UDG formation and high-spin haloes and highlighting that mergers are not necessary to explain the extended sizes in field UDGs. 

Comparing the timing for the last major merger (see Fig.~\ref{fig:histo_lmm} in Appendix~\ref{app:mergers}) with the time where the spin-up of UDG haloes happen, around $t\sim 4$-$6$ Gyr, a casual link between both events seems unsupported, casting doubts on the last major merger as culprit of the high $\lambda$. In fact, this is in agreement with the idea that mergers only temporarily increase the spins of haloes, with any excess spin disappearing once the particles with the largest angular momentum move outside of the virialized region \citep{Donghia2007}. For instance, UDGs in the {\sc Romulus25} simulation, where early mergers are believed to play a role, also show an instantaneous spin increase but no excess spin in the sample at $z=0$. This is different from our results, where UDGs have a large $\lambda$ parameter at $z=0$ suggesting that the formation scenario for our sample is different than that in \citet{Wright2021}.

\subsection{Star formation, color and stellar age}
\label{ssec:color}

Like other galaxies, UDGs in observations show a clear bimodality in their stellar populations when comparing different environments: they are red, quiescent and old in high-density regions \citep[][]{vanderBurg2016, Lee2020, FerreMateu2018} while they are star-forming, blue and gas-bearing in the field  \citep[e.g., ][]{He2019,Jackson2021,Rong2020b,Kadowaki2021}. Reproducing these trends is important for any theoretical model of UDG formation, a benchmark that is attainable in our sample thanks to the large volume of the TNG50 simulation. 

\begin{figure*}
	\centering
	\includegraphics[width=0.86\textwidth]{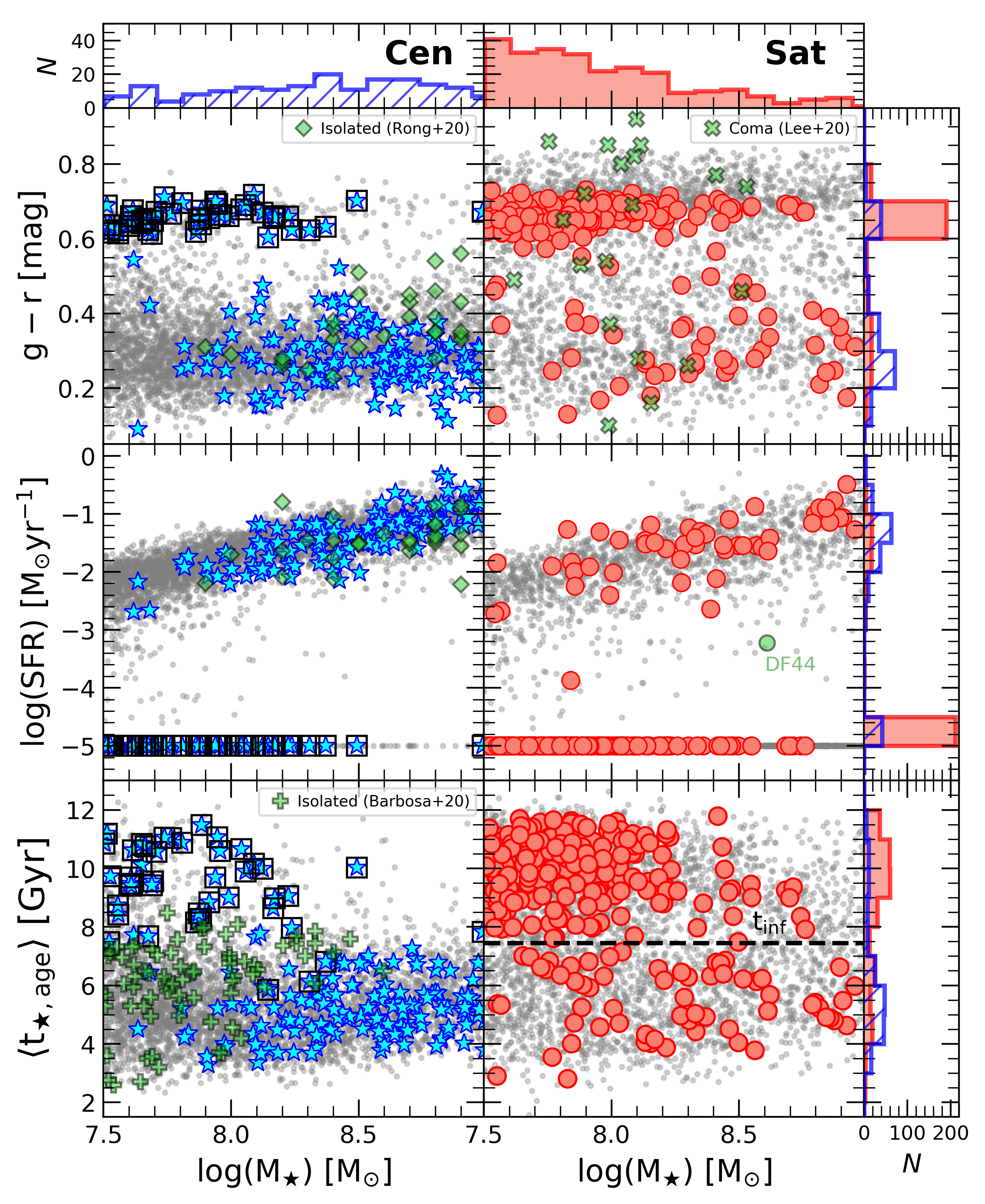}
	\caption{Stellar population properties of simulated UDGs in different environments as a function of stellar mass. From top to bottom: $g$-$r$ color, star formation rate and median stellar ages. Galaxies with zero SFR are artificially placed at log(SFR)$=-5$ for plotting purposes. Central UDGs (left column) are bluer (top), typically star-forming and with relatively young stellar ages (backsplash UDGs highlighted on black squares) while satellites are red, quiescent and older, with typical median stellar ages $<t_\star> \sim 10$ Gyr.  The dashed black line in the stellar age for the satellites panel corresponds to the median of infall times of $\rm{ t_{inf} \sim 7.5 ~ Gyr}$ ago. Histograms along both axes show the distributions of central and satellite UDGs in each quantity. When possible, we compare with available observations as quoted in each panel (green symbols). Simulated UDGs in all environments follow the trends found in observations.}
\label{fig:features}
\end{figure*}

Fig.~\ref{fig:features} shows the color (top), star-formation (middle) and average stellar age (bottom) for our simulated UDGs in the field (blue stars, left column) and satellites (red circles, right column). Non-UDG dwarfs in this mass range are shown, as before, in grey. For comparison, data from observations is added, when available, and highlighted in green symbols \citep{Rong2020b, Lee2020, LeeChris2020, Barbosa2020}. Additionally, we have checked that field UDGs have rich gas reservoirs, with $\rm{M_{gas} \sim 7.5 \times 10^8 ~ M_{\odot}}$ on average within twice the half mass radius of the stars, in agreement with observations of gas-rich UDGs in the field \citep{Leisman2017,Spekkens_and_Karunakaran2018}. On average, field UDGs (non-backsplash) have larger gas fractions than non-UDG centrals ($\rm{M_{gas}/M_\star ~2}$ and $0.66$, respectively). We find a good agreement between theoretical predictions and the properties of observed UDGs in different environments, providing support for the realism of the properties predicted for UDGs in TNG50. 

While the majority of UDGs follow the general expectations described above, there are a handful of objects that behave differently. In the field population, there is a subsample of red, old and quiescent UDGs, which were shown to be backsplash objects in previous work \citep{Benavides2021}. Interestingly, while only about $5\%$ of field UDGs show these characteristics, the fraction increases as we consider lower stellar masses, representing about $25\%$ of field UDGs for dwarfs with $\rm{M_\star} \sim 10^{7.5}~$\msun\; \citep[see ][ for a detailed discussion]{Benavides2021}. On the other hand, there is a small number of satellite UDGs that populate the blue cloud, show non-negligible star formation and younger ages ($5.21 \pm 1.11$ Gyr). We have checked that these correspond to objects with recent infall times ($\lesssim 2$ Gyr ago) and, encouragingly, these kind of objects seem to also be present in observational samples judging by, for example, intermediate color UDGs in Coma \citep{Lee2020} or color-gradients with cluster-centric distance and overall environment \citep{Kadowaki2021}. 

\subsection{Morphology}
\label{ssec:morpho}

The intrinsic morphology and shapes of UDGs may place important constraints on their formation mechanism \citep{Burkert2017}. Following \citet{Sales2012}, we quantify morphology by means of $\kappa_{\rm rot}$, a ratio that compares the energy in rotational support to the total kinetic energy of the stellar particles in a galaxy. More specifically,
\begin{equation}
\kappa_{\rm rot} = \frac{K_{rot}}{K}=\frac{1}{K} \sum\;  \frac{1}{2} m \left( \frac{j_z}{R} \right)^2
\label{eq:krot}	\ \ ,
\end{equation}
where $j_z$ is the z-component of the angular momentum of each stellar particle so that the direction of the total angular momentum of the galaxy is on the z-axis, $m$ is their mass, $R$ is their cylindrical radii and the sum is over stars within the galaxy radius. Large values for $\kappa_{\rm rot} \geq 0.7$ are associated with rotationally supported systems, or disks, while dispersion dominated objects with $\kappa_{\rm rot} \leq 0.35$ are more associated with traditional bulges. Intermediate values appear with galaxies that have coexisting bulge and disk components, or dynamically hotter disks supported partially by dispersion. 

\begin{figure*}
		\includegraphics[width=\columnwidth]{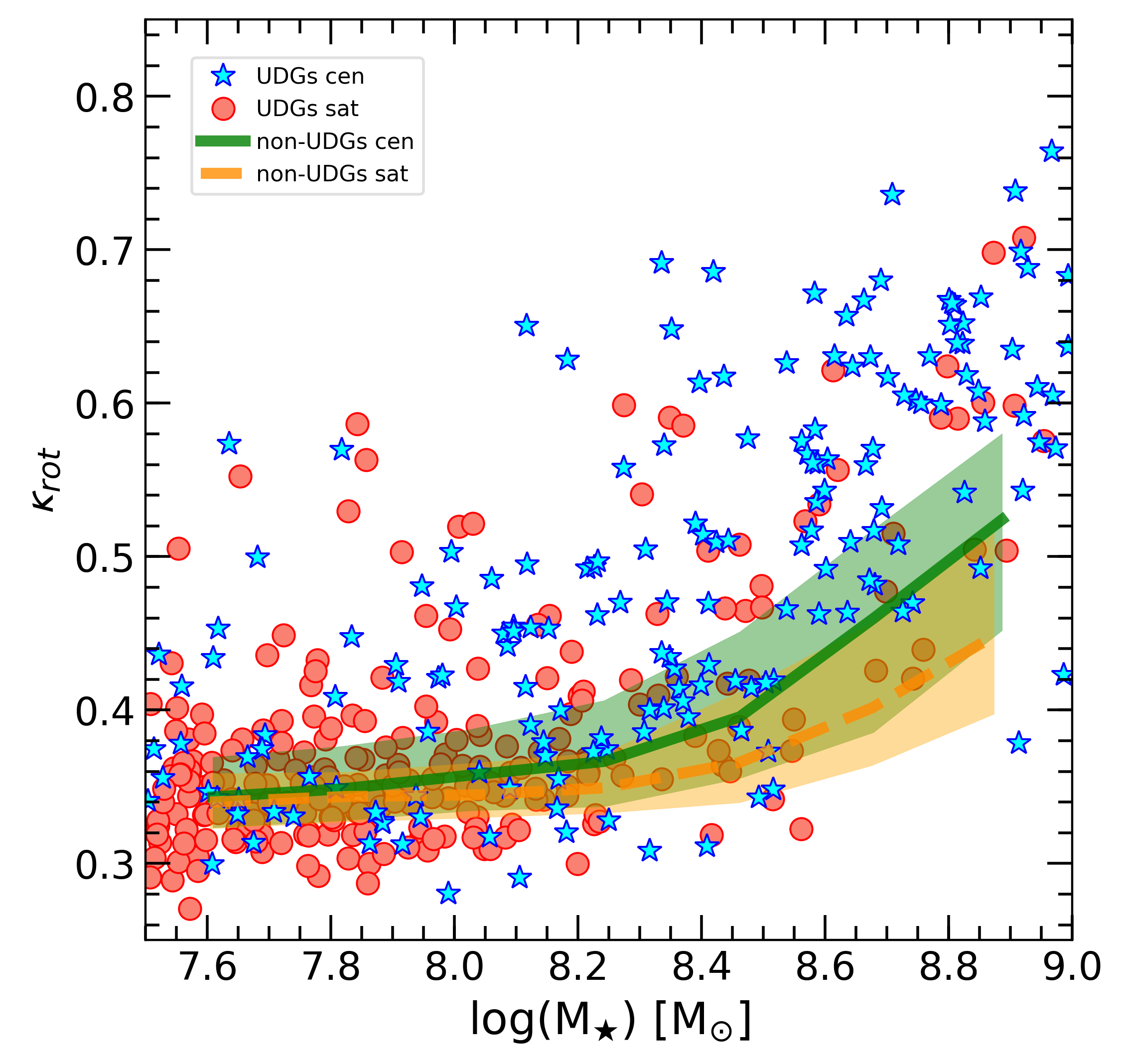}
		\includegraphics[width=\columnwidth]{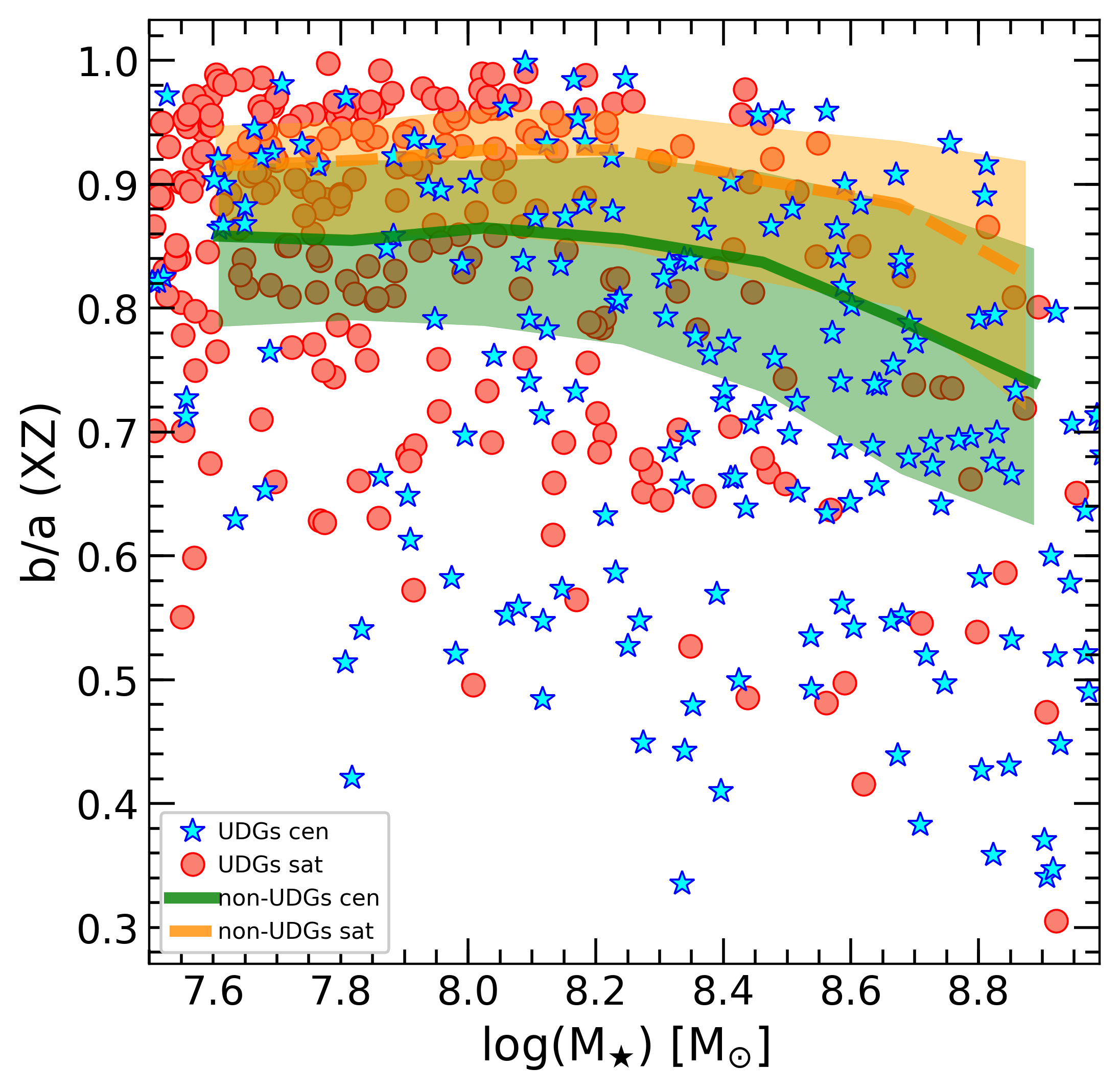}
		\caption{{\it Left:} Morphology \citep[as measured by $\kappa_{rot}$,][]{Sales2012} as a function of stellar mass. Large $k_{rot}$ values are indicative of rotational support and therefore of disk-dominated morphologies, while low values are representative of spheroidal objects. More massive simulated UDGs tend to be disky while low mass counterparts are more dispersion dominated, with little difference between the central (blue stars) and satellite (red circles) populations. The median and 25$^{\rm th}$-75$^{\rm th}$ percentiles of non-UDG dwarfs are shown by the solid lines and shaded areas, which show similar trends to the UDG samples. {\it Right:} another indicator of morphology, quantifying shapes by the projected axis ratio $q=b/a$ measured at $r=2r_{h,\star}$. Simulated UDGs show a relatively flat axis ratio distribution, in agreement with observations, with a tail extending to lower $q$ values on the more massive end due to the presence of more disky morphologies.}
	\label{fig:morpho}
\end{figure*}

Fig.~\ref{fig:morpho} (left panel) shows the distribution of $\kappa_{\rm rot}$ as a function of stellar mass for our sample. We find a wide range of intrinsic morphologies and rotational support, in agreement with other simulation results \citep{CardonaBarrero2020}. This is interesting since the formation mechanism proposed for the UDGs in the NIHAO simulations are dominated by powerful outflows and not necessarily correlated with halo spin \citep{DiCintio2017}. Morphologies might have some constraining power to distinguish fundamentally different formation scenarios (internal vs. external for example), but might not be strict enough to pinpoint exactly which of the internal processes is dominant.  

Two interesting points arise from the morphologies predicted for UDGs in TNG50. First, rotational support is common in the more massive UDGs ($\rm{ M_\star \geq 10^{8.5} }$~\msun) where one might expect to see disky morphologies, but they are mostly dispersion-dominated in the low mass end of our sample, a feature that also holds for non-UDG dwarfs in TNG50 (see green and orange shading in Fig.~\ref{fig:morpho}). Second, there are no marked differences in the morphology of central and satellite UDGs, suggesting that environmental transformations for the satellites act more quickly on star-formation indicators than on morphology, in agreement with previous results in observations and simulations \citep{Joshi2021,Roman2017b,Kadowaki2021}.

Similar trends are spotted when looking at the projected shapes of simulated UDGs (right panel Fig.~\ref{fig:morpho}), a good alternative to morphology for observational samples. 2D shapes are measured for random projections by using the normalized inertia tensor with all stellar particles within twice the half-mass radius. The specific radius used to measure shapes should not in principle impact the results too much, as ellipticity has been shown to be relatively independent of surface brightness in low surface brightness samples \citep{Kado-Fong2021}. 

Overall, our sample displays a wide distribution of axis ratios, with $<q>=<b/a> \sim 0.78 \pm 0.17$ and typical values between $q \sim 0.4$ and $q=1$. This compares well with, for example, measurements of UDGs in low-density environments by \citet{Roman2017b}, which report  $<q> \sim 0.67 \pm 0.13$, compared to the value for our central UDGs $<q> \sim 0.73 \pm 0.18$, and very similar to the value $<b/a> = 0.72$ reported by \citet{ManceraPina2019a}.

At a fixed stellar mass, central or satellite UDGs display similar shapes, in agreement with our conclusions on morphology. Note that the model predicts a noticeable shift from large $q \sim 0.9$ values in the low mass end of our sample to a much more uniform distribution of $q=[0.4$-$1]$ at the high mass end, corresponding to the shift from dispersion-dominated objects in fainter UDGs to rotation-dominated disky galaxies for the most massive UDGs. 

Results from observations also suggest a relatively flat distribution of axis ratios  \citep{Koda2015,Rong2020,Kado-Fong2021}, which are interpreted in favour of oblate intrinsic shapes for observed UDGs \citep[however see ][for a different view]{Burkert2017}. Our results, in particular for $\rm{M_\star \geq 10^8}$~\msun\; agree with that picture. 

It is puzzling, however, that UDGs in our sample show an excess of halo spin independent of galaxy mass, but the shapes and morphologies do show a strong dependence with $\rm{M_\star}$. A combination of feedback strength and the particular ISM model in TNG50 are the likely culprits for this kinematic transition from disky dwarfs to dispersion-dominated in the low mass end, which has also been highlighted in other codes and for non-UDG galaxies \citep{Wheeler2017,Carlsten2021}. It would be interesting to explore whether this mass (or luminosity) dependence on UDG shapes is supported by observations or whether this is a direct result of the particular baryonic modeling implemented in this simulation.

\section{Abundance and structural evolution of satellite UDGs}
\label{sec:sats}

\begin{figure}
\centering
\includegraphics[width=\columnwidth]{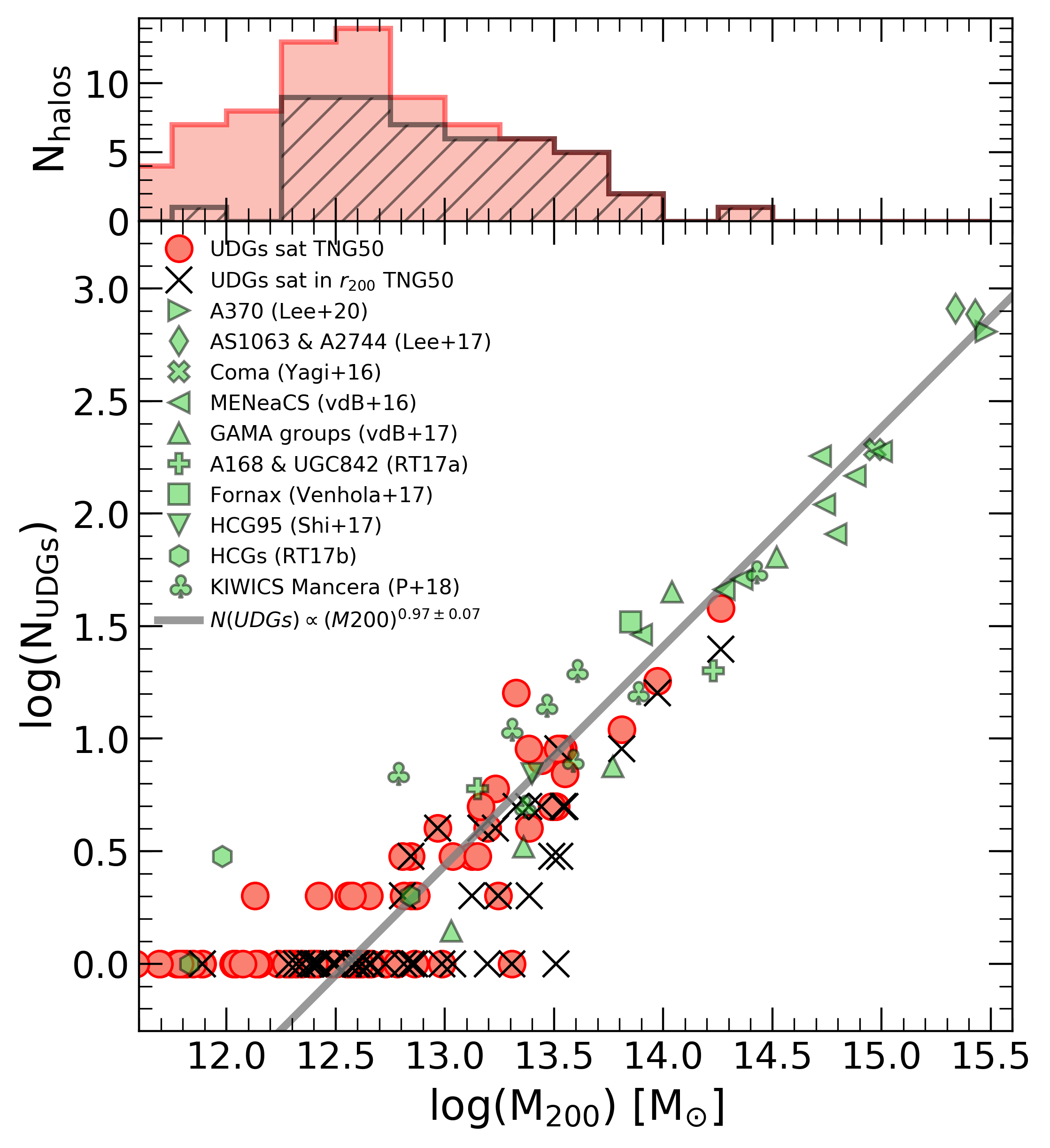}
\caption{Number of UDGs as a function of the virial mass of their host system for the population of satellite UDGs (red circles for all satellites, black crosses for those within $r_{200}$ of the host). Several number of observational data are included in smooth green symbols \citep{Lee2020, Lee2017, Yagi2016, vanderBurg2016, vanderBurg2017, Roman2017a, Venhola2017, Shi2017, Roman2017b, ManceraPina2018}. The grey line indicates the best-fit power-law relation for the simulated UDGs in the host systems with	$\rm{M_{200} > 10^{13} ~M_{\odot}}$ (with slope $n=0.97 \pm 0.07$) which agrees well with observational results of a quasi-linear scaling of UDG abundance with host halo mass.}
\label{fig:abundance}
\end{figure}

An important validation of theoretical models for the formation of UDGs comes from reproducing the observed scaling between the number of UDGs ($N_{\rm UDGs}$) and the host halo mass. Fig.~\ref{fig:abundance} shows in green symbols observational results from several studies that suggest a close-to-linear relation between $N_{\rm UDGs}$ and $\rm{M_{200}}$ in the regime galaxies, groups and low-mass clusters spanned by the TNG50 box. Our simulated UDGs in TNG50 seem to reproduce such a scaling, with slight variations depending on whether satellite UDGs are counted as part of a FoF group (red circles) or within the virial radius (black crosses). 

We note that while reproducing the abundance of UDGs per host halo mass seen in observations is a necessary validation of our sample, a rigorous comparison between theory and observations as well as across observational samples is not feasible, since the selection criteria of UDGs, radial extent of the survey and systematic biases may vary across different studies, all factors that impact the number of reported UDGs \citep{ManceraPina2018,VanNest2022}. Instead, the rough agreement on the normalization and slope of Fig.~\ref{fig:abundance} between simulations and observations indicates that satellite UDGs in TNG50 might be forming with a reasonable frequency, providing a good testbed for studying the role of environment and, in particular, tides in our sample. 

\begin{figure}
\centering
\includegraphics[width=\columnwidth]{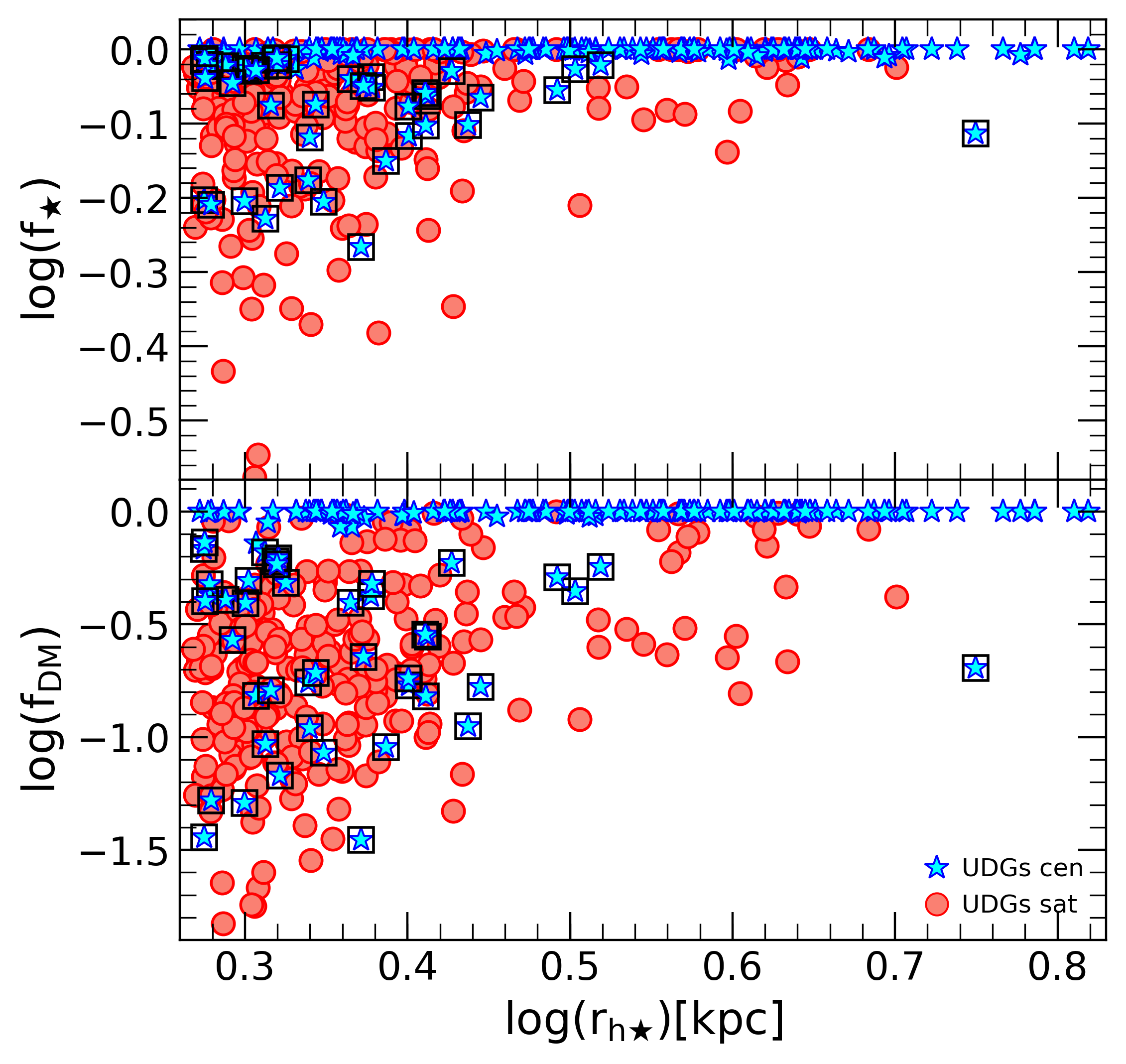}
	\caption{Bound mass fraction for stars (top) and dark matter (bottom) as a function of the stellar size of simulated UDGs. In both cases, the bound fraction is defined as the value at z=0 compared to their maximum value recorded over time, which in the case of the dark matter typically corresponds to infall time. Blue stars and red circles indicate central and satellite UDGs, the black squares highlight the same population of backsplash quenched UDGs from previous figures. Note that satellites UDGs have undergone significant stripping, retaining typically $\approx 20\%$ of their maximum dark matter mass and $\approx 90\%$ of the stellar mass, values independent of size. There are, however, several outliers where tidal stripping has been more pronounced, removing $\geq 50\%$ of the stars.}
\label{fig:fbound}
\end{figure}

We start by quantifying the degree of tidal disruption experienced by satellite UDGs. Fig.~\ref{fig:fbound} shows the fraction of bound stellar ($f_\star$, top) and dark matter ($f_{\rm DM}$, bottom) mass retained for our sample of satellite UDGs (red symbols). Bound fractions are computed by dividing the present-day stellar or dark-matter mass by the maximum mass in either component for each galaxy: 

\begin{equation}
f_{X} = \frac{M^{z=0}_{X}}{M^{\rm{max}}_{X} } .
\label{eq:fbound} 
\end{equation}

\noindent
With this definition, the maximum dark matter mass coincides, in general, with the snapshot prior to infall.

The bottom panel of Fig.~\ref{fig:fbound} indicates that tides  have substantially affected the dark matter content of satellite UDGs, which retain $\sim 20\%$ (median) of their peak dark matter content, with some extreme candidates having lost $ \geq 97\%$ of their dark matter mass. For comparison, we show also the central UDG population in blue which, with the exception of backsplash objects (highlighted in black squares), shows no dark matter depletion, as expected.   

Due to the more centrally concentrated stellar component, tidal disruption is appreciably lower in stellar mass for satellite UDGs (upper panel of Fig.~\ref{fig:fbound}), showing a $\sim 90\%$ (median) bound stellar mass at present-day compared to $\sim 99\%$ in central non-backsplash central UDGs. This means that, in our sample, tidal features in observed UDGs are predicted to be uncommon, which is in good agreement with current observational constraints \citep{Marleau2021}. In individual cases, however, tides might affect more severely the stellar mass, with some of the most extreme satellite UDGs ($\lesssim 5 \%$) retaining only  $\sim 50\%$ or less of their peak stellar mass. 

Cases of satellite UDGs with significant stellar mass loss are rare in TNG50 (for instance, only $\sim 5\%$ of the sample shows $f_\star < 50\%$), but it demonstrates that some surviving satellite UDGs are experiencing severe tidal disruption, in agreement with some early evidence for tidal streams in a few observed UDGs \citep{Toloba2018,Montes2020}. An important caveat to consider is that numerical resolution effects might be accelerating the total tidal disruption of satellites in cosmological simulations \citep{vandenBosch2018,Errani2022}, resulting on artificial disruption of the most extremely tidally affected UDGs. As such, these numbers should be considered as upper limits, especially towards the tail of low bound mass fractions.

\begin{figure}
\centering
\includegraphics[width=\columnwidth]{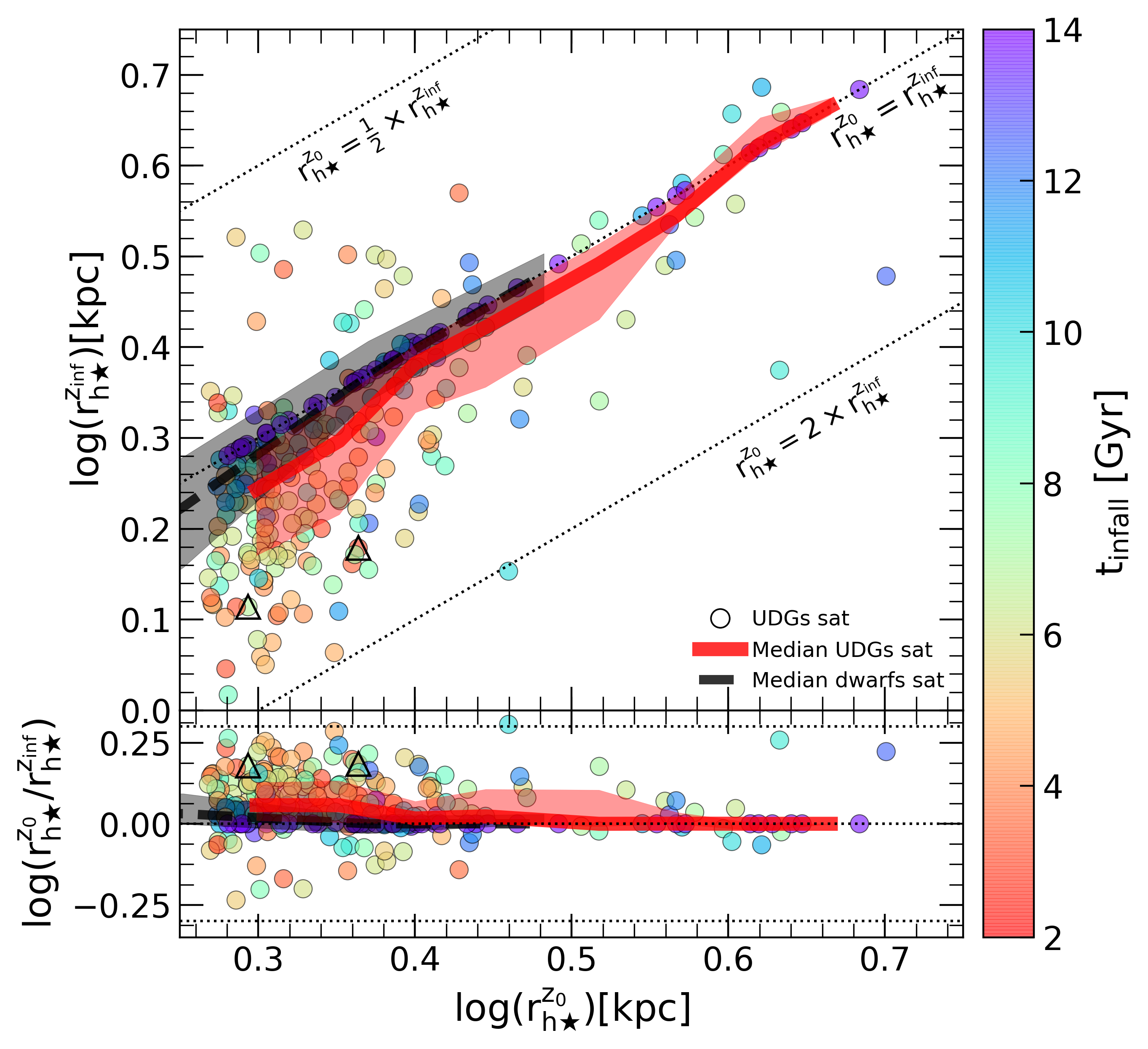}
	\caption{Relationship between the stellar sizes of the satellite UDGs at $z = 0$ vs. at infall time, color-codded by $t_{\rm inf}$. The median infall value at a given present-day size is shown by the thick red line with the shaded region outlining the $25^{\rm th}$ - $75^{\rm th}$ percentiles. UDGs do not significantly change their sizes after infall, with the average expansion value $\sim 16\%$ for the entire sample. For comparison, we also include the median and percentiles for non-UDGs satellites, shown in black. While the overall UDG population is already extended at infall, we do note some outliers, especially in the low mass end, where present-day sizes can be a factor $\geq 1.5$ that of infall, confirming that environmental effects play a role for the formation of some UDGs in our sample. Black open triangles highlight the two examples shown in Fig.~\ref{fig:puffed_examples}.}
\label{fig:rh_infall}
\end{figure}

Tidal effects have been deemed fully \citep[e.g.,][]{Safarzadeh2017, Carleton2019} or partially \citep[e.g.,][]{Jiang2019b, Sales2020, Tremmel2020} responsible for the formation of UDGs in several theoretical models, implying that without the effect of tides, present-day (satellite) UDGs would be normal dwarfs galaxies. We explore this in Fig.~\ref{fig:rh_infall}, which shows for our individual satellite UDGs (colored symbols) a comparison of their stellar half mass radius at infall ($y$-axis) vs. at $z=0$ ($x$-axis). The bottom panel also shows the fractional change as a function of the final size at $z=0$. 

The overwhelming majority of satellite UDGs fall near the 1:1 line in Fig.~\ref{fig:rh_infall}, suggesting little $r_{h,\star}$ evolution due to environmental effects. The median in our UDG sample is shown with the black solid line and shaded areas indicate 25$^{\rm th}$-75$^{\rm th}$ percentiles. For comparison, we also show the median relation between infall and present-day stellar size for non-UDG satellites (grey dotted line). While UDGs do experience a slightly larger size growth, it is only a modest change: satellite UDGs experience a $10\%$ (median) size increase since infall, which is even smaller for the most extended dwarfs and for those with late infall times (color coding). Satellite UDGs were, in general, already extended in the field prior to infall. 

Notwithstanding, tides do play a significant role in at least {\it some} of our satellite UDGs: about $10\%$ of our sample show a stellar half-mass radius increase larger than $50\%$ and would probably not been classified as UDGs without this size evolution within the host environment. These tend to be (although not exclusively) lower mass dwarfs and earlier infalls. Within this group with significant size increase, we have identified two kinds of behaviour: $1)$ rapid stellar expansion associated with infall and quenching and $2)$ a more secular expansion lasting from infall until today; with approximately the sample dividing half-half between these categories.

\begin{figure*}
\includegraphics[width=\columnwidth]{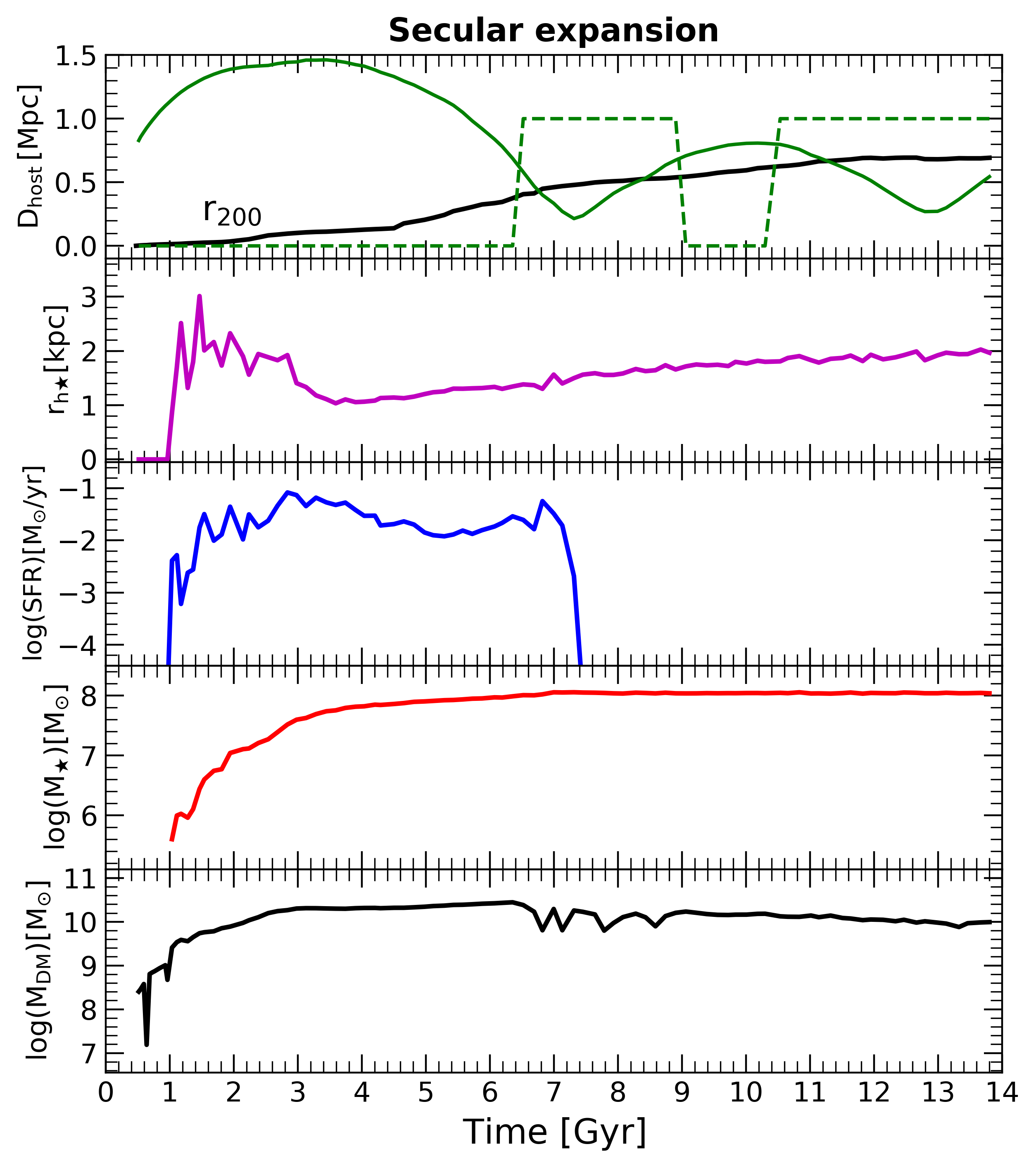}
\includegraphics[width=\columnwidth]{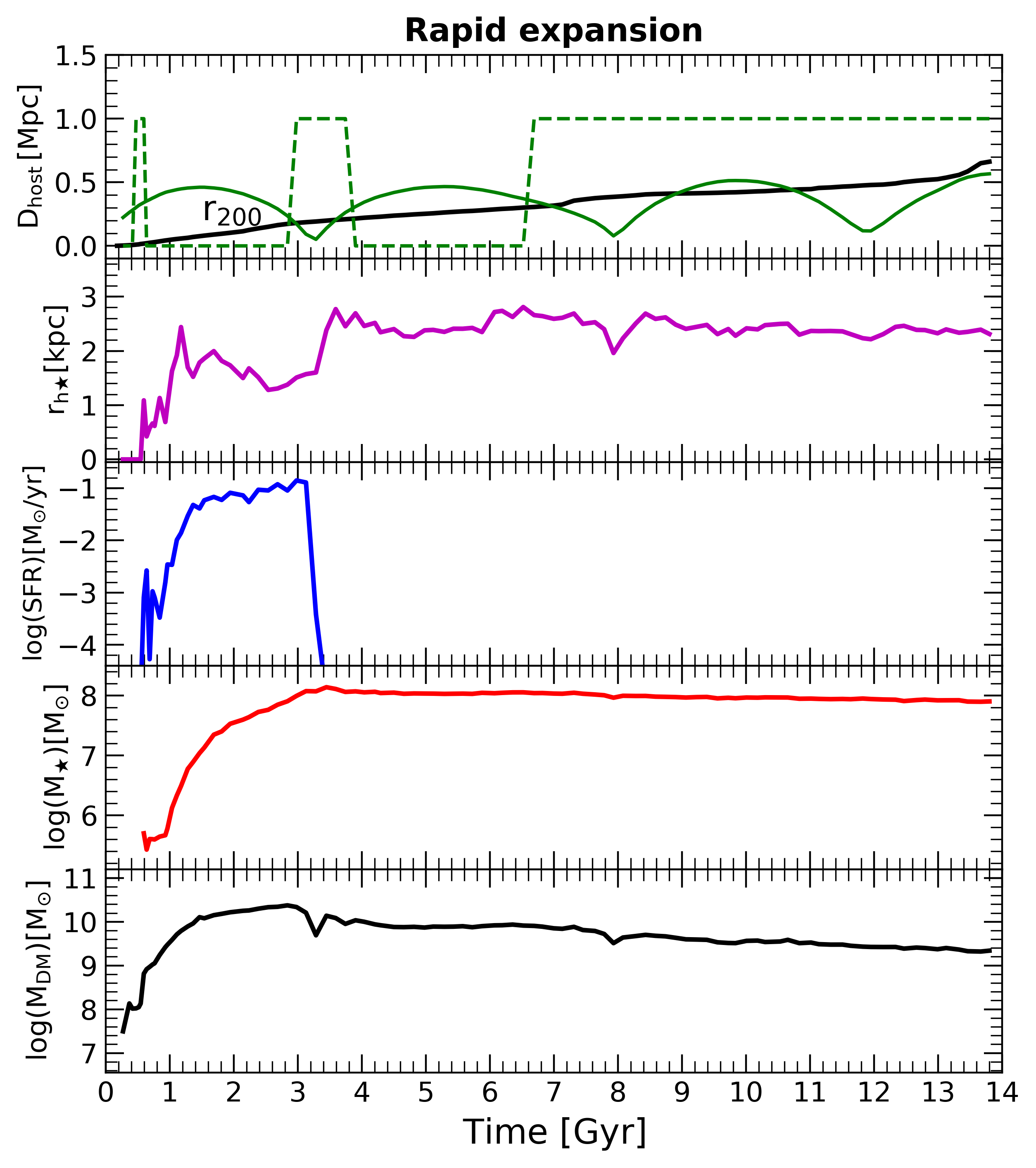}
	\caption{Examples for the evolution of two UDGs with significant size growth after infall ($r_{h \, \star}^{z_0} \geq 1.5 r_{h \, \star}^{z_{\rm inf}}$), highlighted with open triangles in Fig.~\ref{fig:rh_infall}. From the top to bottom: $i)$ cluster-centric distance in the green continuous line while the black line indicates the growth of the virial radius of the host halo, and the green dash line indicates the change from central (0) to satellite (1) along the time, $ii)$ evolution of the stellar half mass radius, $iii)$ star-formation rate, $iv)$ stellar mass of the galaxy (inside of $2 \, r_{h \, \star}$) and $v)$ halo dark matter content. The left column shows an example where the stellar half-mass radius follows a more secular (slow) growth while the right column displays an example where the stellar size doubles quickly after the first pericenter passage. Of the minority of UDGs with significant size change after infall, we find $\sim$half with slow evolution and the other half with rapid evolution akin to these examples in the left and right columns, respectively.
	}
\label{fig:puffed_examples}
\end{figure*}

For illustration, Fig.~\ref{fig:puffed_examples} shows one example of each kind, with a secular stellar radius growth on the left column and a rapid expansion case on the right. Note that for the ``secular expansion" case on the left, quenching occurs at the first pericenter (third row) while $r_{h,\star}$ continues to increase with a slow pace until the present day. This galaxy shows almost no tidal stripping  and suggests that the slow size increase could be a combination of tidal heating and an ageing stellar population in the central regions, although this would warrant a more detailed study of its own given the complexities of disentangling these two intertwined mechanisms. 

An overall numerical resolution effect is unlikely to be responsible for the secular size evolution seen in the example of the left panel of Fig.~\ref{fig:puffed_examples}: non-UDG galaxies of the same mass/size are on average consistent with no-change in $r_{h,\star}$, suggesting that the resolution in the simulation is able to properly handle sizes for similar objects. As a reference, the final half-mass radius increase of this particular UDG candidate shown in the left panel of Fig.~\ref{fig:puffed_examples} is $r_{h,\star}^{z_0} = 1.5 r_{h,\star}^{z_{\rm inf}}$ which is significantly above the median $1.1 r_{h,\star}^{z_{\rm inf}}$ of the whole satellite UDG sample. About $5\%$ of the satellite UDGs show a $\geq 50\%$ size increase with similarly slow time evolution, being largely sub-dominant in the whole satellite UDG population.

On the other hand, the right column of Fig.~\ref{fig:puffed_examples} illustrates the opposite example: a satellite UDG where the size undergoes a ``rapid expansion" event. The increase in stellar size (second row) occurs at the first pericenter around the host group, coincidental in time with the removal of gas and quenching of this galaxy. Such objects (about $\sim 5\%$ of the satellite UDGs) are consistent with being normal dwarfs ``puffed up" by impulsive tidal effects such as non-adiabatic removal of gas \citep{Safarzadeh2017} or tidal heating \citep{Jiang2019}, both mechanisms proposed by previous theoretical models of UDG formation. 

This particular galaxy depicted on the right column of Fig.~\ref{fig:puffed_examples} has also experienced substantial tidal disruption in both, stellar and dark matter components (bound fractions $f_\star =0.58$ and $f_{\rm DM}=0.09$, respectively), but the time-locality of the size increase and its conjunction with the pericenter suggest that the slower tidal stripping of the collisionless components did not drive the net size increase in these kind of objects. 

The predicted satellite UDG population is therefore composed of two types of objects, those born UDG in the field and later accreted into the host haloes and a minority ($\sim 10\%$) where tidal effects cause an appreciable expansion of the stellar distribution. Tidal stripping affects substantially the dark matter and stellar content of all satellite UDGs, having lost $\sim 80\%$ and $\sim 10\%$ on average, respectively, of their peak mass. Our results suggest a scenario where the extended sizes of UDGs are set in their majority due to internal processes before infall, but external environmentally-driven mechanisms play a non-negligible role in transforming some normal dwarfs into UDGs. These results agree well with previous works that proposed a combination of internal and externally-driven effects to explain the satellite UDG population \citep{Jiang2019, Sales2020}, as far as we highlight that only about $10\%$ of satellite UDGs owe their extended sizes to the influence of environment. More specifically, $\sim 10\%$ experience a size increase larger than $50\%$ their infall value, while the median for the  whole satellite UDGs population is only an increase $\sim 16\%$ and were therefore already extended before infall onto their respective hosts.

\section{Discussion}
\label{sec:disc}

Some of the results presented in this work may be at odds with findings reported previously using different numerical simulations. For instance, high halo spins are not needed in formation mechanisms associated with powerful outflows \citep{DiCintio2017,Chan2018}, while other teams have reported no biases in the spin distribution of UDGs and non-UDG objects \citep[e.g., ][]{Jiang2019,Wright2021}. These differences are not unexpected, as the properties and morphologies of simulated galaxies have been shown to depend strongly on the particular feedback prescription implemented \citep{Sales2010, Scannapieco2012}. For the TNG50 baryonic physics treatment, high halo spin seem to play a major role on setting galaxy sizes, at least in the regime of dwarf galaxies explored here. A constructive path forward to compare different theoretical models is to identify a set of predictions that might be used in the near future to validate this particular UDG formation path proposed here. We briefly discuss three of them here: kinematics, number of UDGs and tidal features.

We start with considerations on the kinematics of UDGs, a topic that we defer for a detailed study in forthcoming work (Doppel et al., {\it in-prep}). Observational studies have found a wide range of dark matter content in UDGs \citep{Toloba2018, Doppel2021, Gannon2022} with many suggesting dwarf-size haloes \citep[e.g., ][]{Papastergis2017,Jones2018}, which would be in rough agreement with our results. There are, however, individual objects with peculiar kinematics that are difficult to reconcile with most of the formation scenarios for UDGs, such as the case of dark-matter free galaxies like DF2 and DF4 \citep{vanDokkum2018,Wasserman2018,Danieli2019,vanDokkum2019a_DF4}, or the suggestion from gas-rich field UDGs to have a much lower than expected dark matter mass \citep{ManceraPina2019b,ManceraPina2022}. 

The formation scenario proposed here, together with the baryonic implementation in TNG50 resulting in no dark matter core formation, suggest that such dark-matter poor objects will be difficult to reproduce in our sample or even completely absent \citep[in fact, see ][ for a quantitative discussion]{Kong2022}. However, it is still early times in the observational studies of UDG kinematics and the division between trends for the general UDG population versus the existence of outliers or rare galaxies is currently unclear. In addition, misalignment between gas and stars may complicate the inclination corrections for the rotation curves of gar-rich field UDGs \citep{Gault2021}. Once more observational data becomes available to constrain the dynamical masses of field and satellite UDGs, the internal kinematics of these galaxies will represent a solid validation tool for theoretical models. Note that the power of studies such as the one presented here in TNG50 relies on reproducing population trends, and not individual objects, which might require specific initial or boundary conditions to reproduce specific traits. 

The abundance of UDGs as a function of environment, in particular in the field, is a promising avenue to constrain UDG formation models. The abundance of satellite UDGs is also important, but their interpretation is complicated by membership assignment and distance to the host considerations, among others. In the field, for instance, \citet{Jones2018} measured the abundance of H{\sc i}-bearing UDGs in the ALFALFA survey and determined a cosmic number density of $(1.5 \pm 0.6) \times 10^{-3} ~ {\rm Mpc}^{-3}$, a value found to be too small compared to predictions from semi-analytical models in \citet{Rong2017} where large spins were directly linked to UDG formation. We have checked that the abundance of central UDGs in our sample is $\sim 1.41 \times  10^{-3} ~ {\rm Mpc}^{-3}$, in reasonable agreement with estimates from ALFALFA. 

While a more careful comparison to observational determinations of UDG cosmic abundance is warranted, in particular bearing in mind the effects of different definitions \citep{ManceraPina2018,VanNest2022}, there seems to be no direct evidence indicating that the formation frequency of central (gas-rich) UDGs in TNG50 is too large, despite its link to large spin values. As we highlighted in Sec.~\ref{ssec:spin}, haloes with large $\lambda$ in TNG50 will preferentially form extended galaxies that might qualify as UDGs, but not all high-spin haloes host a central UDG in our simulations, with halo mass and fraction of angular momentum retained also playing a role. The predicted frequency of formation of field UDGs might be completely different in models where mergers or outflows are involved, making observational studies aimed at constraining the abundance of field (and satellite) UDGs a promising tool to help constrain theoretical models. 

Lastly, we argue here that determining the frequency of tidal streams associated to UDGs is of extreme importance. In our model, while tides are responsible for the transformation of a low fraction of normal dwarfs into UDGs, the large majority of UDGs in TNG50 owe their extended sizes to internal halo properties. We therefore expect a relatively low incidence of stellar streams around observed UDGs in high-density environments and not at all for UDGs in the field. Observations of UDGs in low and intermediate-density environments seem to agree with this picture \citep{Marleau2021}, but more studies are needed. This may become one of the most important predictions to be confirmed in the near future,  as more observational campaigns become increasingly capable of surveying the extremely low-surface brightness universe.

\section{Summary}
\label{sec:concl}

We use the TNG50 hydrodynamical cosmological simulation to study the formation of UDGs in the stellar mass range $\rm{M_\star} = 10^{7.5}$-$10^{9}~$\msun. The large volume simulated in TNG50 allows for one of the first self-consistent studies of the formation of UDGs in different environments, spanning from the field to galaxy clusters with virial mass $\rm{M_{200} \sim 10^{14}}~$\msun. We define UDGs as outliers in the mass-size relation, selecting at a given stellar mass, the $5\%$ of objects with the most extended stellar half-mass radii $r_{h,\star}$. Such a selection retrieves a group of low surface brightness galaxies ($\Sigma \sim [24.5$-$28]$ mag arcsec$^{-2}$ measured within the effective radius and assuming mass-to-light ratio $1$\; and $r_{h,\star} \geq 2$ kpc) that are in good agreement with common selection criteria of UDGs in observations. Our sample consists of 176 central (or field) UDGs and 260 satellite UDGs inhabiting host haloes with virial masses $\rm{M_{200}} \sim 10^{12}$-$10^{14.3}~$\msun. 

The main result in this paper is that the large majority of UDGs (both, centrals and satellites) form in TNG50 due to internal processes, in particular, due to dark matter haloes with a high spin. The median halo spin for the central UDG sample is $< \lambda > = 0.059$ compared to $< \lambda > = 0.035$ for the non-UDG sample. Satellites also show an excess spin when measured at the time of infall ($< \lambda > = 0.047$). This result is in agreement with one of the first theoretical explanations for the extended sizes of UDGs using semi-analytical models \citep{Amorisco2016}, being now confirmed using hydrodynamical simulations. For example, \citet{Amorisco2016}  predict that satellite UDGs should have typical median values $\rm \lambda \sim [0.040, 0.063]$, which agrees well with the median in our simulated sample of central and satellite UDGs, $\lambda \sim 0.06, 0.04$, respectively.

Our UDG sample inhabits dwarf-mass haloes like other non-UDGs in the same stellar mass range, with $\rm{M_{200}}=[10^{10} \rm - 10^{11}]~$\msun. Within this range, simulated UDGs are biased-high in $\rm{M_{200}}$, having masses $\sim 40\%$ - $70\%$ higher than non-UDG at a fixed $\rm{M_\star}$. A third factor seems to be determining the large stellar sizes in simulated (central) UDGs. In addition to living in haloes with large spins, their stellar components manage to capture $\sim 3$ times higher specific angular momentum fraction from the halo than non-UDG dwarfs. These three factors (high spin, biased-high mass or virial radius and large angular momentum retention) are common ingredients in analytical disk formation models, such as \citet{mmw1998}, and seem to explain well the formation of UDGs as dwarfs in the extreme tail-end of extended stellar sizes.  

In good agreement with observations, simulated UDGs in TNG50 are blue, young and star-forming in stark contrast with satellite UDGs being red, old and quiescent. There are exceptions to these main features: red and quiescent UDGs can be found in the field in significant numbers due to backsplash orbits \citep[analyzed in detail in ][]{Benavides2021} and a small fraction of blue star-forming satellite UDGs might result from recent infall times in the outskirts of groups and clusters. Environmental effects act quickly to stop star formation in satellite UDGs, with quiescent satellite UDGs having typical median stellar ages $t_{\star,\rm age} \sim 10$ Gyr. 

We find that the extended sizes of most satellite UDGs  are not the result of tidal evolution in the groups and clusters but instead were already in place before infall. For instance, only $10\%$ of satellite UDGs show a $\geq 50\%$ increase in $r_{h,\star}$ compared to infall. Most satellite UDGs were therefore already UDGs in the field before accretion, and the environment is responsible for their quiescence and ageing of their stellar population, but does not play a dominant role in setting their extended sizes. A small but significant fraction of satellite UDGs form due to external or environmentally-driven processes.  

For the $\sim 10\%$ of UDGs showing a  $\geq 50\%$  increase in size since infall, we find that tidal effects such as sudden gas removal and tidal heating at pericenter passages are responsible for triggering the size increase, which can be rapid (about half of the cases) or secular (for the remaining half). A combination of internal (dominant) and external factors seem necessary to explain the population of satellite UDGs in TNG50, in agreement with previous theoretical results \citep{Jiang2019,Sales2020}. Tidal effects are important for the mass content of satellite UDGs, which lose on average $\sim 80\%$ of their peak dark matter mass and $\sim 10\%$ of their stars.  

We argue in Sec.~\ref{sec:disc} that internal kinematics, abundance of UDGs as a function of environment and  the presence of (or lack of) tidal streams around UDGs are among the most promising paths to further constrain the formation of UDGs in theoretical models.

\section*{Acknowledgments}
The authors would like to thank Annalisa Pillepich and the whole TNG50 team for early access to the simulation data. JAB and MGA acknowledge financial support from CONICET through PIP 11220170100527CO grant. LVS is grateful for partial financial support from UC-Mexus, NASA-ATP-80NSSC20K0566 and NSF-CAREER-1945310 grants. 

\section*{Data Availability}

This paper is based on halo catalogs and merger trees from the Illustris-TNG Project \citep{Nelson2019TNG, Nelson2019TNG50}. These data are publicly available at \href{https://www.tng-project.org/}{https://www.tng-project.org/}. The main properties of the UDG and non-UDG dwarf galaxy samples, and other products included in this analysis, may be shared upon request to the corresponding author if no further conflict exists with ongoing projects. 



\bibliographystyle{mnras}
\bibliography{ms_UDGs.bib} 

\begin{thebibliography}{}
\makeatletter
\relax
\def\mn@urlcharsother{\let\do\@makeother \do\$\do\&\do\#\do\^\do\_\do\%\do\~}
\def\mn@doi{\begingroup\mn@urlcharsother \@ifnextchar [ {\mn@doi@}
  {\mn@doi@[]}}
\def\mn@doi@[#1]#2{\def\@tempa{#1}\ifx\@tempa\@empty \href
  {http://dx.doi.org/#2} {doi:#2}\else \href {http://dx.doi.org/#2} {#1}\fi
  \endgroup}
\def\mn@eprint#1#2{\mn@eprint@#1:#2::\@nil}
\def\mn@eprint@arXiv#1{\href {http://arxiv.org/abs/#1} {{\tt arXiv:#1}}}
\def\mn@eprint@dblp#1{\href {http://dblp.uni-trier.de/rec/bibtex/#1.xml}
  {dblp:#1}}
\def\mn@eprint@#1:#2:#3:#4\@nil{\def\@tempa {#1}\def\@tempb {#2}\def\@tempc
  {#3}\ifx \@tempc \@empty \let \@tempc \@tempb \let \@tempb \@tempa \fi \ifx
  \@tempb \@empty \def\@tempb {arXiv}\fi \@ifundefined
  {mn@eprint@\@tempb}{\@tempb:\@tempc}{\expandafter \expandafter \csname
  mn@eprint@\@tempb\endcsname \expandafter{\@tempc}}}

\bibitem[\protect\citeauthoryear{{Amorisco} \& {Loeb}}{{Amorisco} \&
  {Loeb}}{2016}]{Amorisco2016}
{Amorisco} N.~C.,  {Loeb} A.,  2016, \mn@doi [\mnras] {10.1093/mnrasl/slw055},
  \href {https://ui.adsabs.harvard.edu/abs/2016MNRAS.459L..51A} {459, L51}

\bibitem[\protect\citeauthoryear{{Amorisco}, {Monachesi}, {Agnello}  \&
  {White}}{{Amorisco} et~al.}{2018}]{Amorisco2018}
{Amorisco} N.~C.,  {Monachesi} A.,  {Agnello} A.,   {White} S.~D.~M.,  2018,
  \mn@doi [\mnras] {10.1093/mnras/sty116}, \href
  {https://ui.adsabs.harvard.edu/abs/2018MNRAS.475.4235A} {475, 4235}

\bibitem[\protect\citeauthoryear{{Barbosa} et~al.,}{{Barbosa}
  et~al.}{2020}]{Barbosa2020}
{Barbosa} C.~E.,  et~al., 2020, \mn@doi [\apjs] {10.3847/1538-4365/ab7660},
  \href {https://ui.adsabs.harvard.edu/abs/2020ApJS..247...46B} {247, 46}

\bibitem[\protect\citeauthoryear{{Beasley}, {Romanowsky}, {Pota}, {Navarro},
  {Martinez Delgado}, {Neyer}  \& {Deich}}{{Beasley}
  et~al.}{2016}]{Beasley2016}
{Beasley} M.~A.,  {Romanowsky} A.~J.,  {Pota} V.,  {Navarro} I.~M.,  {Martinez
  Delgado} D.,  {Neyer} F.,   {Deich} A.~L.,  2016, \mn@doi [\apjl]
  {10.3847/2041-8205/819/2/L20}, \href
  {https://ui.adsabs.harvard.edu/abs/2016ApJ...819L..20B} {819, L20}

\bibitem[\protect\citeauthoryear{{Behroozi}, {Wechsler}  \&
  {Conroy}}{{Behroozi} et~al.}{2013}]{Behroozi2013}
{Behroozi} P.~S.,  {Wechsler} R.~H.,   {Conroy} C.,  2013, \mn@doi [\apj]
  {10.1088/0004-637X/770/1/57}, \href
  {https://ui.adsabs.harvard.edu/abs/2013ApJ...770...57B} {770, 57}

\bibitem[\protect\citeauthoryear{{Benavides}, {Sales}  \& {Abadi}}{{Benavides}
  et~al.}{2020}]{Benavides2020}
{Benavides} J.~A.,  {Sales} L.~V.,   {Abadi} M.~G.,  2020, \mn@doi [\mnras]
  {10.1093/mnras/staa2636}, \href
  {https://ui.adsabs.harvard.edu/abs/2020MNRAS.498.3852B} {498, 3852}

\bibitem[\protect\citeauthoryear{{Benavides} et~al.,}{{Benavides}
  et~al.}{2021}]{Benavides2021}
{Benavides} J.~A.,  et~al., 2021, \mn@doi [Nature Astronomy]
  {10.1038/s41550-021-01458-1}, \href
  {https://ui.adsabs.harvard.edu/abs/2021NatAs...5.1255B} {5, 1255}

\bibitem[\protect\citeauthoryear{{Brook} et~al.,}{{Brook}
  et~al.}{2011}]{Brook2011}
{Brook} C.~B.,  et~al., 2011, \mn@doi [\mnras]
  {10.1111/j.1365-2966.2011.18545.x}, \href
  {https://ui.adsabs.harvard.edu/abs/2011MNRAS.415.1051B} {415, 1051}

\bibitem[\protect\citeauthoryear{{Brook}, {Stinson}, {Gibson}, {Ro{\v{s}}kar},
  {Wadsley}  \& {Quinn}}{{Brook} et~al.}{2012}]{Brook2012}
{Brook} C.~B.,  {Stinson} G.,  {Gibson} B.~K.,  {Ro{\v{s}}kar} R.,  {Wadsley}
  J.,   {Quinn} T.,  2012, \mn@doi [\mnras] {10.1111/j.1365-2966.2011.19740.x},
  \href {https://ui.adsabs.harvard.edu/abs/2012MNRAS.419..771B} {419, 771}

\bibitem[\protect\citeauthoryear{{Bullock}, {Dekel}, {Kolatt}, {Kravtsov},
  {Klypin}, {Porciani}  \& {Primack}}{{Bullock} et~al.}{2001}]{Bullock2001}
{Bullock} J.~S.,  {Dekel} A.,  {Kolatt} T.~S.,  {Kravtsov} A.~V.,  {Klypin}
  A.~A.,  {Porciani} C.,   {Primack} J.~R.,  2001, \mn@doi [\apj]
  {10.1086/321477}, \href
  {https://ui.adsabs.harvard.edu/abs/2001ApJ...555..240B} {555, 240}

\bibitem[\protect\citeauthoryear{{Burkert}}{{Burkert}}{2017}]{Burkert2017}
{Burkert} A.,  2017, \mn@doi [\apj] {10.3847/1538-4357/aa671c}, \href
  {https://ui.adsabs.harvard.edu/abs/2017ApJ...838...93B} {838, 93}

\bibitem[\protect\citeauthoryear{{Cardona-Barrero}, {Di Cintio}, {Brook},
  {Ruiz-Lara}, {Beasley}, {Falc{\'o}n-Barroso}  \&
  {Macci{\`o}}}{{Cardona-Barrero} et~al.}{2020}]{CardonaBarrero2020}
{Cardona-Barrero} S.,  {Di Cintio} A.,  {Brook} C. B.~A.,  {Ruiz-Lara} T.,
  {Beasley} M.~A.,  {Falc{\'o}n-Barroso} J.,   {Macci{\`o}} A.~V.,  2020,
  \mn@doi [\mnras] {10.1093/mnras/staa2094}, \href
  {https://ui.adsabs.harvard.edu/abs/2020MNRAS.497.4282C} {497, 4282}

\bibitem[\protect\citeauthoryear{{Carleton}, {Errani}, {Cooper}, {Kaplinghat},
  {Pe{\~n}arrubia}  \& {Guo}}{{Carleton} et~al.}{2019}]{Carleton2019}
{Carleton} T.,  {Errani} R.,  {Cooper} M.,  {Kaplinghat} M.,  {Pe{\~n}arrubia}
  J.,   {Guo} Y.,  2019, \mn@doi [\mnras] {10.1093/mnras/stz383}, \href
  {https://ui.adsabs.harvard.edu/abs/2019MNRAS.485..382C} {485, 382}

\bibitem[\protect\citeauthoryear{{Carleton}, {Guo}, {Munshi}, {Tremmel}  \&
  {Wright}}{{Carleton} et~al.}{2021}]{Carleton2021}
{Carleton} T.,  {Guo} Y.,  {Munshi} F.,  {Tremmel} M.,   {Wright} A.,  2021,
  \mn@doi [\mnras] {10.1093/mnras/stab031}, \href
  {https://ui.adsabs.harvard.edu/abs/2021MNRAS.502..398C} {502, 398}

\bibitem[\protect\citeauthoryear{{Carlsten}, {Greene}, {Greco}, {Beaton}  \&
  {Kado-Fong}}{{Carlsten} et~al.}{2021}]{Carlsten2021}
{Carlsten} S.~G.,  {Greene} J.~E.,  {Greco} J.~P.,  {Beaton} R.~L.,
  {Kado-Fong} E.,  2021, \mn@doi [\apj] {10.3847/1538-4357/ac2581}, \href
  {https://ui.adsabs.harvard.edu/abs/2021ApJ...922..267C} {922, 267}

\bibitem[\protect\citeauthoryear{{Chabrier}}{{Chabrier}}{2003}]{Chabrier2003}
{Chabrier} G.,  2003, \mn@doi [\pasp] {10.1086/376392}, \href
  {https://ui.adsabs.harvard.edu/abs/2003PASP..115..763C} {115, 763}

\bibitem[\protect\citeauthoryear{{Chan}, {Kere{\v{s}}}, {Wetzel}, {Hopkins},
  {Faucher-Gigu{\`e}re}, {El-Badry}, {Garrison-Kimmel}  \&
  {Boylan-Kolchin}}{{Chan} et~al.}{2018}]{Chan2018}
{Chan} T.~K.,  {Kere{\v{s}}} D.,  {Wetzel} A.,  {Hopkins} P.~F.,
  {Faucher-Gigu{\`e}re} C.~A.,  {El-Badry} K.,  {Garrison-Kimmel} S.,
  {Boylan-Kolchin} M.,  2018, \mn@doi [\mnras] {10.1093/mnras/sty1153}, \href
  {https://ui.adsabs.harvard.edu/abs/2018MNRAS.478..906C} {478, 906}

\bibitem[\protect\citeauthoryear{{Conselice}}{{Conselice}}{2018}]{Conselice2018}
{Conselice} C.~J.,  2018, \mn@doi [Research Notes of the American Astronomical
  Society] {10.3847/2515-5172/aab7f6}, \href
  {https://ui.adsabs.harvard.edu/abs/2018RNAAS...2...43C} {2, 43}

\bibitem[\protect\citeauthoryear{{D'Onghia} \& {Navarro}}{{D'Onghia} \&
  {Navarro}}{2007}]{Donghia2007}
{D'Onghia} E.,  {Navarro} J.~F.,  2007, \mn@doi [\mnras]
  {10.1111/j.1745-3933.2007.00348.x}, \href
  {https://ui.adsabs.harvard.edu/abs/2007MNRAS.380L..58D} {380, L58}

\bibitem[\protect\citeauthoryear{{Dalcanton}, {Spergel}, {Gunn}, {Schmidt}  \&
  {Schneider}}{{Dalcanton} et~al.}{1997}]{Dalcanton1997}
{Dalcanton} J.~J.,  {Spergel} D.~N.,  {Gunn} J.~E.,  {Schmidt} M.,
  {Schneider} D.~P.,  1997, \mn@doi [\aj] {10.1086/118499}, \href
  {https://ui.adsabs.harvard.edu/abs/1997AJ....114..635D} {114, 635}

\bibitem[\protect\citeauthoryear{{Danieli}, {van Dokkum}, {Conroy}, {Abraham}
  \& {Romanowsky}}{{Danieli} et~al.}{2019}]{Danieli2019}
{Danieli} S.,  {van Dokkum} P.,  {Conroy} C.,  {Abraham} R.,   {Romanowsky}
  A.~J.,  2019, \mn@doi [\apjl] {10.3847/2041-8213/ab0e8c}, \href
  {https://ui.adsabs.harvard.edu/abs/2019ApJ...874L..12D} {874, L12}

\bibitem[\protect\citeauthoryear{{Danieli} et~al.,}{{Danieli}
  et~al.}{2022}]{Danieli2022}
{Danieli} S.,  et~al., 2022, \mn@doi [\apjl] {10.3847/2041-8213/ac590a}, \href
  {https://ui.adsabs.harvard.edu/abs/2022ApJ...927L..28D} {927, L28}

\bibitem[\protect\citeauthoryear{{Davis}, {Efstathiou}, {Frenk}  \&
  {White}}{{Davis} et~al.}{1985}]{Davis1985}
{Davis} M.,  {Efstathiou} G.,  {Frenk} C.~S.,   {White} S.~D.~M.,  1985,
  \mn@doi [\apj] {10.1086/163168}, \href
  {https://ui.adsabs.harvard.edu/abs/1985ApJ...292..371D} {292, 371}

\bibitem[\protect\citeauthoryear{{Di Cintio}, {Brook}, {Dutton}, {Macci{\`o}},
  {Obreja}  \& {Dekel}}{{Di Cintio} et~al.}{2017}]{DiCintio2017}
{Di Cintio} A.,  {Brook} C.~B.,  {Dutton} A.~A.,  {Macci{\`o}} A.~V.,  {Obreja}
  A.,   {Dekel} A.,  2017, \mn@doi [\mnras] {10.1093/mnrasl/slw210}, \href
  {https://ui.adsabs.harvard.edu/abs/2017MNRAS.466L...1D} {466, L1}

\bibitem[\protect\citeauthoryear{{Dolag}, {Borgani}, {Murante}  \&
  {Springel}}{{Dolag} et~al.}{2009}]{Dolag2009}
{Dolag} K.,  {Borgani} S.,  {Murante} G.,   {Springel} V.,  2009, \mn@doi
  [\mnras] {10.1111/j.1365-2966.2009.15034.x}, \href
  {https://ui.adsabs.harvard.edu/abs/2009MNRAS.399..497D} {399, 497}

\bibitem[\protect\citeauthoryear{{Doppel}, {Sales}, {Navarro}, {Abadi}, {Peng},
  {Toloba}  \& {Ramos-Almendares}}{{Doppel} et~al.}{2021}]{Doppel2021}
{Doppel} J.~E.,  {Sales} L.~V.,  {Navarro} J.~F.,  {Abadi} M.~G.,  {Peng}
  E.~W.,  {Toloba} E.,   {Ramos-Almendares} F.,  2021, \mn@doi [\mnras]
  {10.1093/mnras/staa3915}, \href
  {https://ui.adsabs.harvard.edu/abs/2021MNRAS.502.1661D} {502, 1661}

\bibitem[\protect\citeauthoryear{{Doroshkevich}}{{Doroshkevich}}{1970}]{Doroshkevich1970}
{Doroshkevich} A.~G.,  1970, Astrofizika, \href
  {https://ui.adsabs.harvard.edu/abs/1970Afz.....6..581D} {6, 581}

\bibitem[\protect\citeauthoryear{{Dutton} \& {van den Bosch}}{{Dutton} \& {van
  den Bosch}}{2012}]{Dutton2012}
{Dutton} A.~A.,  {van den Bosch} F.~C.,  2012, \mn@doi [\mnras]
  {10.1111/j.1365-2966.2011.20339.x}, \href
  {https://ui.adsabs.harvard.edu/abs/2012MNRAS.421..608D} {421, 608}

\bibitem[\protect\citeauthoryear{{Errani}, {Navarro}, {Ibata}  \&
  {Pe{\~n}arrubia}}{{Errani} et~al.}{2022}]{Errani2022}
{Errani} R.,  {Navarro} J.~F.,  {Ibata} R.,   {Pe{\~n}arrubia} J.,  2022,
  \mn@doi [\mnras] {10.1093/mnras/stac476}, \href
  {https://ui.adsabs.harvard.edu/abs/2022MNRAS.511.6001E} {511, 6001}

\bibitem[\protect\citeauthoryear{{Ferr{\'e}-Mateu} et~al.,}{{Ferr{\'e}-Mateu}
  et~al.}{2018}]{FerreMateu2018}
{Ferr{\'e}-Mateu} A.,  et~al., 2018, \mn@doi [\mnras] {10.1093/mnras/sty1597},
  \href {https://ui.adsabs.harvard.edu/abs/2018MNRAS.479.4891F} {479, 4891}

\bibitem[\protect\citeauthoryear{{Gannon} et~al.,}{{Gannon}
  et~al.}{2022}]{Gannon2022}
{Gannon} J.~S.,  et~al., 2022, \mn@doi [\mnras] {10.1093/mnras/stab3297}, \href
  {https://ui.adsabs.harvard.edu/abs/2022MNRAS.510..946G} {510, 946}

\bibitem[\protect\citeauthoryear{{Gault} et~al.,}{{Gault}
  et~al.}{2021}]{Gault2021}
{Gault} L.,  et~al., 2021, \mn@doi [\apj] {10.3847/1538-4357/abd79d}, \href
  {https://ui.adsabs.harvard.edu/abs/2021ApJ...909...19G} {909, 19}

\bibitem[\protect\citeauthoryear{{Guo}, {White}, {Li}  \&
  {Boylan-Kolchin}}{{Guo} et~al.}{2010}]{Guo2010}
{Guo} Q.,  {White} S.,  {Li} C.,   {Boylan-Kolchin} M.,  2010, \mn@doi [\mnras]
  {10.1111/j.1365-2966.2010.16341.x}, \href
  {https://ui.adsabs.harvard.edu/abs/2010MNRAS.404.1111G} {404, 1111}

\bibitem[\protect\citeauthoryear{{Harris}, {Harris}  \& {Hudson}}{{Harris}
  et~al.}{2015}]{Harris2015}
{Harris} W.~E.,  {Harris} G.~L.,   {Hudson} M.~J.,  2015, \mn@doi [\apj]
  {10.1088/0004-637X/806/1/36}, \href
  {https://ui.adsabs.harvard.edu/abs/2015ApJ...806...36H} {806, 36}

\bibitem[\protect\citeauthoryear{{Harris}, {Blakeslee}  \& {Harris}}{{Harris}
  et~al.}{2017}]{Harris2017}
{Harris} W.~E.,  {Blakeslee} J.~P.,   {Harris} G. L.~H.,  2017, \mn@doi [\apj]
  {10.3847/1538-4357/836/1/67}, \href
  {https://ui.adsabs.harvard.edu/abs/2017ApJ...836...67H} {836, 67}

\bibitem[\protect\citeauthoryear{{He}, {Wu}, {Du}, {Wicker}, {Zhao}, {Lei}  \&
  {Liu}}{{He} et~al.}{2019}]{He2019}
{He} M.,  {Wu} H.,  {Du} W.,  {Wicker} J.,  {Zhao} P.,  {Lei} F.,   {Liu} J.,
  2019, \mn@doi [\apj] {10.3847/1538-4357/ab2710}, \href
  {https://ui.adsabs.harvard.edu/abs/2019ApJ...880...30H} {880, 30}

\bibitem[\protect\citeauthoryear{{Hernquist}}{{Hernquist}}{1990}]{Hernquist1990}
{Hernquist} L.,  1990, \mn@doi [\apj] {10.1086/168845}, \href
  {https://ui.adsabs.harvard.edu/abs/1990ApJ...356..359H} {356, 359}

\bibitem[\protect\citeauthoryear{{Hopkins}, {Cox}, {Younger}  \&
  {Hernquist}}{{Hopkins} et~al.}{2009}]{Hopkins2009}
{Hopkins} P.~F.,  {Cox} T.~J.,  {Younger} J.~D.,   {Hernquist} L.,  2009,
  \mn@doi [\apj] {10.1088/0004-637X/691/2/1168}, \href
  {https://ui.adsabs.harvard.edu/abs/2009ApJ...691.1168H} {691, 1168}

\bibitem[\protect\citeauthoryear{{Impey}, {Bothun}  \& {Malin}}{{Impey}
  et~al.}{1988}]{Impey1988}
{Impey} C.,  {Bothun} G.,   {Malin} D.,  1988, \mn@doi [\apj] {10.1086/166500},
  \href {https://ui.adsabs.harvard.edu/abs/1988ApJ...330..634I} {330, 634}

\bibitem[\protect\citeauthoryear{{Jackson} et~al.,}{{Jackson}
  et~al.}{2021}]{Jackson2021}
{Jackson} R.~A.,  et~al., 2021, \mn@doi [\mnras] {10.1093/mnras/stab077}, \href
  {https://ui.adsabs.harvard.edu/abs/2021MNRAS.502.4262J} {502, 4262}

\bibitem[\protect\citeauthoryear{{Jiang}, {Dekel}  \& {Freundlich}}{{Jiang}
  et~al.}{2019a}]{Jiang2019}
{Jiang} F.,  {Dekel} A.,   {Freundlich} J.,  2019a, in {Di Matteo} P.,
  {Creevey} O.,  {Crida} A.,  {Kordopatis} G.,  {Malzac} J.,  {Marquette}
  J.~B.,  {N'Diaye} M.,   {Venot} O.,  eds, SF2A-2019: Proceedings of the
  Annual meeting of the French Society of Astronomy and Astrophysics. p.~Di

\bibitem[\protect\citeauthoryear{{Jiang}, {Dekel}, {Freundlich}, {Romanowsky},
  {Dutton}, {Macci{\`o}}  \& {Di Cintio}}{{Jiang} et~al.}{2019b}]{Jiang2019b}
{Jiang} F.,  {Dekel} A.,  {Freundlich} J.,  {Romanowsky} A.~J.,  {Dutton}
  A.~A.,  {Macci{\`o}} A.~V.,   {Di Cintio} A.,  2019b, \mn@doi [\mnras]
  {10.1093/mnras/stz1499}, \href
  {https://ui.adsabs.harvard.edu/abs/2019MNRAS.487.5272J} {487, 5272}

\bibitem[\protect\citeauthoryear{{Jones}, {Papastergis}, {Pandya}, {Leisman},
  {Romanowsky}, {Yung}, {Somerville}  \& {Adams}}{{Jones}
  et~al.}{2018}]{Jones2018}
{Jones} M.~G.,  {Papastergis} E.,  {Pandya} V.,  {Leisman} L.,  {Romanowsky}
  A.~J.,  {Yung} L.~Y.~A.,  {Somerville} R.~S.,   {Adams} E.~A.~K.,  2018,
  \mn@doi [\aap] {10.1051/0004-6361/201732409}, \href
  {https://ui.adsabs.harvard.edu/abs/2018A&A...614A..21J} {614, A21}

\bibitem[\protect\citeauthoryear{{Jones}, {Bennet}, {Mutlu-Pakdil}, {Sand},
  {Spekkens}, {Crnojevic}, {Karunakaran}  \& {Zaritsky}}{{Jones}
  et~al.}{2021}]{Jones2021}
{Jones} M.~G.,  {Bennet} P.,  {Mutlu-Pakdil} B.,  {Sand} D.~J.,  {Spekkens} K.,
   {Crnojevic} D.,  {Karunakaran} A.,   {Zaritsky} D.,  2021, arXiv e-prints,
  \href {https://ui.adsabs.harvard.edu/abs/2021arXiv210412805J} {p.
  arXiv:2104.12805}

\bibitem[\protect\citeauthoryear{{Joshi}, {Pillepich}, {Nelson}, {Zinger},
  {Marinacci}, {Springel}, {Vogelsberger}  \& {Hernquist}}{{Joshi}
  et~al.}{2021}]{Joshi2021}
{Joshi} G.~D.,  {Pillepich} A.,  {Nelson} D.,  {Zinger} E.,  {Marinacci} F.,
  {Springel} V.,  {Vogelsberger} M.,   {Hernquist} L.,  2021, \mn@doi [\mnras]
  {10.1093/mnras/stab2573}, \href
  {https://ui.adsabs.harvard.edu/abs/2021MNRAS.508.1652J} {508, 1652}

\bibitem[\protect\citeauthoryear{{Kado-Fong} et~al.,}{{Kado-Fong}
  et~al.}{2021}]{Kado-Fong2021}
{Kado-Fong} E.,  et~al., 2021, \mn@doi [\apj] {10.3847/1538-4357/ac15f0}, \href
  {https://ui.adsabs.harvard.edu/abs/2021ApJ...920...72K} {920, 72}

\bibitem[\protect\citeauthoryear{{Kadowaki}, {Zaritsky}, {Donnerstein}, {RS},
  {Karunakaran}  \& {Spekkens}}{{Kadowaki} et~al.}{2021}]{Kadowaki2021}
{Kadowaki} J.,  {Zaritsky} D.,  {Donnerstein} R.~L.,  {RS} P.,  {Karunakaran}
  A.,   {Spekkens} K.,  2021, \mn@doi [\apj] {10.3847/1538-4357/ac2948}, \href
  {https://ui.adsabs.harvard.edu/abs/2021ApJ...923..257K} {923, 257}

\bibitem[\protect\citeauthoryear{{Koda}, {Yagi}, {Yamanoi}  \&
  {Komiyama}}{{Koda} et~al.}{2015}]{Koda2015}
{Koda} J.,  {Yagi} M.,  {Yamanoi} H.,   {Komiyama} Y.,  2015, \mn@doi [\apjl]
  {10.1088/2041-8205/807/1/L2}, \href
  {https://ui.adsabs.harvard.edu/abs/2015ApJ...807L...2K} {807, L2}

\bibitem[\protect\citeauthoryear{{Kong}, {Kaplinghat}, {Yu}, {Fraternali}  \&
  {Mancera Pi{\~n}a}}{{Kong} et~al.}{2022}]{Kong2022}
{Kong} D.,  {Kaplinghat} M.,  {Yu} H.-B.,  {Fraternali} F.,   {Mancera
  Pi{\~n}a} P.~E.,  2022, arXiv e-prints, \href
  {https://ui.adsabs.harvard.edu/abs/2022arXiv220405981K} {p. arXiv:2204.05981}

\bibitem[\protect\citeauthoryear{{La Marca} et~al.,}{{La Marca}
  et~al.}{2022}]{LaMarca2022}
{La Marca} A.,  et~al., 2022, arXiv e-prints, \href
  {https://ui.adsabs.harvard.edu/abs/2022arXiv220607385L} {p. arXiv:2206.07385}

\bibitem[\protect\citeauthoryear{{Lee}, {Kang}, {Lee}  \& {Jang}}{{Lee}
  et~al.}{2017}]{Lee2017}
{Lee} M.~G.,  {Kang} J.,  {Lee} J.~H.,   {Jang} I.~S.,  2017, \mn@doi [\apj]
  {10.3847/1538-4357/aa78fb}, \href
  {https://ui.adsabs.harvard.edu/abs/2017ApJ...844..157L} {844, 157}

\bibitem[\protect\citeauthoryear{{Lee}, {Hodges-Kluck}  \& {Gallo}}{{Lee}
  et~al.}{2020a}]{LeeChris2020}
{Lee} C.~H.,  {Hodges-Kluck} E.,   {Gallo} E.,  2020a, \mn@doi [\mnras]
  {10.1093/mnras/staa1955}, \href
  {https://ui.adsabs.harvard.edu/abs/2020MNRAS.497.2759L} {497, 2759}

\bibitem[\protect\citeauthoryear{{Lee}, {Kang}, {Lee}  \& {Jang}}{{Lee}
  et~al.}{2020b}]{Lee2020}
{Lee} J.~H.,  {Kang} J.,  {Lee} M.~G.,   {Jang} I.~S.,  2020b, \mn@doi [\apj]
  {10.3847/1538-4357/ab8632}, \href
  {https://ui.adsabs.harvard.edu/abs/2020ApJ...894...75L} {894, 75}

\bibitem[\protect\citeauthoryear{{Leisman} et~al.,}{{Leisman}
  et~al.}{2017}]{Leisman2017}
{Leisman} L.,  et~al., 2017, \mn@doi [\apj] {10.3847/1538-4357/aa7575}, \href
  {https://ui.adsabs.harvard.edu/abs/2017ApJ...842..133L} {842, 133}

\bibitem[\protect\citeauthoryear{{Lim}, {Peng}, {C{\^o}t{\'e}}, {Sales}, {den
  Brok}, {Blakeslee}  \& {Guhathakurta}}{{Lim} et~al.}{2018}]{Lim2018}
{Lim} S.,  {Peng} E.~W.,  {C{\^o}t{\'e}} P.,  {Sales} L.~V.,  {den Brok} M.,
  {Blakeslee} J.~P.,   {Guhathakurta} P.,  2018, \mn@doi [\apj]
  {10.3847/1538-4357/aacb81}, \href
  {https://ui.adsabs.harvard.edu/abs/2018ApJ...862...82L} {862, 82}

\bibitem[\protect\citeauthoryear{{Lim} et~al.,}{{Lim} et~al.}{2020}]{Lim2020}
{Lim} S.,  et~al., 2020, \mn@doi [\apj] {10.3847/1538-4357/aba433}, \href
  {https://ui.adsabs.harvard.edu/abs/2020ApJ...899...69L} {899, 69}

\bibitem[\protect\citeauthoryear{{Macci{\`o}}, {Dutton}, {van den Bosch},
  {Moore}, {Potter}  \& {Stadel}}{{Macci{\`o}} et~al.}{2007}]{Maccio2007}
{Macci{\`o}} A.~V.,  {Dutton} A.~A.,  {van den Bosch} F.~C.,  {Moore} B.,
  {Potter} D.,   {Stadel} J.,  2007, \mn@doi [\mnras]
  {10.1111/j.1365-2966.2007.11720.x}, \href
  {https://ui.adsabs.harvard.edu/abs/2007MNRAS.378...55M} {378, 55}

\bibitem[\protect\citeauthoryear{{Macci{\`o}}, {Prats}, {Dixon}, {Buck},
  {Waterval}, {Arora}, {Courteau}  \& {Kang}}{{Macci{\`o}}
  et~al.}{2021}]{Maccio2021}
{Macci{\`o}} A.~V.,  {Prats} D.~H.,  {Dixon} K.~L.,  {Buck} T.,  {Waterval} S.,
   {Arora} N.,  {Courteau} S.,   {Kang} X.,  2021, \mn@doi [\mnras]
  {10.1093/mnras/staa3716}, \href
  {https://ui.adsabs.harvard.edu/abs/2021MNRAS.501..693M} {501, 693}

\bibitem[\protect\citeauthoryear{{Mancera Pi{\~n}a}, {Peletier}, {Aguerri},
  {Venhola}, {Trager}  \& {Choque Challapa}}{{Mancera Pi{\~n}a}
  et~al.}{2018}]{ManceraPina2018}
{Mancera Pi{\~n}a} P.~E.,  {Peletier} R.~F.,  {Aguerri} J.~A.~L.,  {Venhola}
  A.,  {Trager} S.,   {Choque Challapa} N.,  2018, \mn@doi [\mnras]
  {10.1093/mnras/sty2574}, \href
  {https://ui.adsabs.harvard.edu/abs/2018MNRAS.481.4381M} {481, 4381}

\bibitem[\protect\citeauthoryear{{Mancera Pi{\~n}a}, {Aguerri}, {Peletier},
  {Venhola}, {Trager}  \& {Choque Challapa}}{{Mancera Pi{\~n}a}
  et~al.}{2019a}]{ManceraPina2019a}
{Mancera Pi{\~n}a} P.~E.,  {Aguerri} J.~A.~L.,  {Peletier} R.~F.,  {Venhola}
  A.,  {Trager} S.,   {Choque Challapa} N.,  2019a, \mn@doi [\mnras]
  {10.1093/mnras/stz238}, \href
  {https://ui.adsabs.harvard.edu/abs/2019MNRAS.485.1036M} {485, 1036}

\bibitem[\protect\citeauthoryear{{Mancera Pi{\~n}a} et~al.,}{{Mancera Pi{\~n}a}
  et~al.}{2019b}]{ManceraPina2019b}
{Mancera Pi{\~n}a} P.~E.,  et~al., 2019b, \mn@doi [\apjl]
  {10.3847/2041-8213/ab40c7}, \href
  {https://ui.adsabs.harvard.edu/abs/2019ApJ...883L..33M} {883, L33}

\bibitem[\protect\citeauthoryear{{Mancera Pi{\~n}a} et~al.,}{{Mancera Pi{\~n}a}
  et~al.}{2020}]{ManceraPina2020}
{Mancera Pi{\~n}a} P.~E.,  et~al., 2020, \mn@doi [\mnras]
  {10.1093/mnras/staa1256}, \href
  {https://ui.adsabs.harvard.edu/abs/2020MNRAS.495.3636M} {495, 3636}

\bibitem[\protect\citeauthoryear{{Mancera Pi{\~n}a}, {Fraternali}, {Oosterloo},
  {Adams}, {Oman}  \& {Leisman}}{{Mancera Pi{\~n}a}
  et~al.}{2022a}]{ManceraPina2022}
{Mancera Pi{\~n}a} P.~E.,  {Fraternali} F.,  {Oosterloo} T.,  {Adams} E. A.~K.,
   {Oman} K.~A.,   {Leisman} L.,  2022a, \mn@doi [\mnras]
  {10.1093/mnras/stab3491}, \href
  {https://ui.adsabs.harvard.edu/abs/2022MNRAS.512.3230M} {512, 3230}

\bibitem[\protect\citeauthoryear{{Mancera Pi{\~n}a}, {Fraternali}, {Oosterloo},
  {Adams}, {di Teodoro}, {Bacchini}  \& {Iorio}}{{Mancera Pi{\~n}a}
  et~al.}{2022b}]{ManceraPina2022b}
{Mancera Pi{\~n}a} P.~E.,  {Fraternali} F.,  {Oosterloo} T.,  {Adams} E. A.~K.,
   {di Teodoro} E.,  {Bacchini} C.,   {Iorio} G.,  2022b, \mn@doi [\mnras]
  {10.1093/mnras/stac1508}, \href
  {https://ui.adsabs.harvard.edu/abs/2022MNRAS.514.3329M} {514, 3329}

\bibitem[\protect\citeauthoryear{{Marinacci} et~al.,}{{Marinacci}
  et~al.}{2018}]{Marinacci2018}
{Marinacci} F.,  et~al., 2018, \mn@doi [\mnras] {10.1093/mnras/sty2206}, \href
  {https://ui.adsabs.harvard.edu/abs/2018MNRAS.480.5113M} {480, 5113}

\bibitem[\protect\citeauthoryear{{Marleau} et~al.,}{{Marleau}
  et~al.}{2021}]{Marleau2021}
{Marleau} F.~R.,  et~al., 2021, \mn@doi [\aap] {10.1051/0004-6361/202141432},
  \href {https://ui.adsabs.harvard.edu/abs/2021A&A...654A.105M} {654, A105}

\bibitem[\protect\citeauthoryear{{Mart{\'\i}n-Navarro}
  et~al.,}{{Mart{\'\i}n-Navarro} et~al.}{2019}]{MartinNavarro2019}
{Mart{\'\i}n-Navarro} I.,  et~al., 2019, \mn@doi [\mnras]
  {10.1093/mnras/stz252}, \href
  {https://ui.adsabs.harvard.edu/abs/2019MNRAS.484.3425M} {484, 3425}

\bibitem[\protect\citeauthoryear{{Mart{\'\i}nez-Delgado}
  et~al.,}{{Mart{\'\i}nez-Delgado} et~al.}{2016}]{MartinezDelgado2016}
{Mart{\'\i}nez-Delgado} D.,  et~al., 2016, \mn@doi [\aj]
  {10.3847/0004-6256/151/4/96}, \href
  {https://ui.adsabs.harvard.edu/abs/2016AJ....151...96M} {151, 96}

\bibitem[\protect\citeauthoryear{{McConnachie} et~al.,}{{McConnachie}
  et~al.}{2008}]{McConnachie2008}
{McConnachie} A.~W.,  et~al., 2008, \mn@doi [\apj] {10.1086/591313}, \href
  {https://ui.adsabs.harvard.edu/abs/2008ApJ...688.1009M} {688, 1009}

\bibitem[\protect\citeauthoryear{{Mihos} et~al.,}{{Mihos}
  et~al.}{2015}]{Mihos2015}
{Mihos} J.~C.,  et~al., 2015, \mn@doi [\apjl] {10.1088/2041-8205/809/2/L21},
  \href {https://ui.adsabs.harvard.edu/abs/2015ApJ...809L..21M} {809, L21}

\bibitem[\protect\citeauthoryear{{Mo}, {Mao}  \& {White}}{{Mo}
  et~al.}{1998}]{mmw1998}
{Mo} H.~J.,  {Mao} S.,   {White} S. D.~M.,  1998, \mn@doi [\mnras]
  {10.1046/j.1365-8711.1998.01227.x}, \href
  {https://ui.adsabs.harvard.edu/abs/1998MNRAS.295..319M} {295, 319}

\bibitem[\protect\citeauthoryear{{Montes}, {Infante-Sainz}, {Madrigal-Aguado},
  {Rom{\'a}n}, {Monelli}, {Borlaff}  \& {Trujillo}}{{Montes}
  et~al.}{2020}]{Montes2020}
{Montes} M.,  {Infante-Sainz} R.,  {Madrigal-Aguado} A.,  {Rom{\'a}n} J.,
  {Monelli} M.,  {Borlaff} A.~S.,   {Trujillo} I.,  2020, \mn@doi [\apj]
  {10.3847/1538-4357/abc340}, \href
  {https://ui.adsabs.harvard.edu/abs/2020ApJ...904..114M} {904, 114}

\bibitem[\protect\citeauthoryear{{Moster}, {Naab}  \& {White}}{{Moster}
  et~al.}{2013}]{Moster2013}
{Moster} B.~P.,  {Naab} T.,   {White} S. D.~M.,  2013, \mn@doi [\mnras]
  {10.1093/mnras/sts261}, \href
  {https://ui.adsabs.harvard.edu/abs/2013MNRAS.428.3121M} {428, 3121}

\bibitem[\protect\citeauthoryear{{Naiman} et~al.,}{{Naiman}
  et~al.}{2018}]{Naiman2018}
{Naiman} J.~P.,  et~al., 2018, \mn@doi [\mnras] {10.1093/mnras/sty618}, \href
  {https://ui.adsabs.harvard.edu/abs/2018MNRAS.477.1206N} {477, 1206}

\bibitem[\protect\citeauthoryear{{Nelson} et~al.,}{{Nelson}
  et~al.}{2018}]{Nelson2018}
{Nelson} D.,  et~al., 2018, \mn@doi [\mnras] {10.1093/mnras/stx3040}, \href
  {https://ui.adsabs.harvard.edu/abs/2018MNRAS.475..624N} {475, 624}

\bibitem[\protect\citeauthoryear{{Nelson} et~al.,}{{Nelson}
  et~al.}{2019a}]{Nelson2019TNG}
{Nelson} D.,  et~al., 2019a, \mn@doi [Computational Astrophysics and Cosmology]
  {10.1186/s40668-019-0028-x}, \href
  {https://ui.adsabs.harvard.edu/abs/2019ComAC...6....2N} {6, 2}

\bibitem[\protect\citeauthoryear{{Nelson} et~al.,}{{Nelson}
  et~al.}{2019b}]{Nelson2019TNG50}
{Nelson} D.,  et~al., 2019b, \mn@doi [\mnras] {10.1093/mnras/stz2306}, \href
  {https://ui.adsabs.harvard.edu/abs/2019MNRAS.490.3234N} {490, 3234}

\bibitem[\protect\citeauthoryear{{Oman}, {Navarro}, {Sales}, {Fattahi},
  {Frenk}, {Sawala}, {Schaller}  \& {White}}{{Oman} et~al.}{2016}]{Oman2016}
{Oman} K.~A.,  {Navarro} J.~F.,  {Sales} L.~V.,  {Fattahi} A.,  {Frenk} C.~S.,
  {Sawala} T.,  {Schaller} M.,   {White} S. D.~M.,  2016, \mn@doi [\mnras]
  {10.1093/mnras/stw1251}, \href
  {https://ui.adsabs.harvard.edu/abs/2016MNRAS.460.3610O} {460, 3610}

\bibitem[\protect\citeauthoryear{{Papastergis}, {Adams}  \&
  {Romanowsky}}{{Papastergis} et~al.}{2017}]{Papastergis2017}
{Papastergis} E.,  {Adams} E.~A.~K.,   {Romanowsky} A.~J.,  2017, \mn@doi
  [\aap] {10.1051/0004-6361/201730795}, \href
  {https://ui.adsabs.harvard.edu/abs/2017A&A...601L..10P} {601, L10}

\bibitem[\protect\citeauthoryear{{Peng} \& {Lim}}{{Peng} \&
  {Lim}}{2016}]{Peng2016}
{Peng} E.~W.,  {Lim} S.,  2016, \mn@doi [\apjl] {10.3847/2041-8205/822/2/L31},
  \href {https://ui.adsabs.harvard.edu/abs/2016ApJ...822L..31P} {822, L31}

\bibitem[\protect\citeauthoryear{{Pillepich} et~al.,}{{Pillepich}
  et~al.}{2018a}]{Pillepich2018a}
{Pillepich} A.,  et~al., 2018a, \mn@doi [\mnras] {10.1093/mnras/stx2656}, \href
  {https://ui.adsabs.harvard.edu/abs/2018MNRAS.473.4077P} {473, 4077}

\bibitem[\protect\citeauthoryear{{Pillepich} et~al.,}{{Pillepich}
  et~al.}{2018b}]{Pillepich2018b}
{Pillepich} A.,  et~al., 2018b, \mn@doi [\mnras] {10.1093/mnras/stx3112}, \href
  {https://ui.adsabs.harvard.edu/abs/2018MNRAS.475..648P} {475, 648}

\bibitem[\protect\citeauthoryear{{Pillepich} et~al.,}{{Pillepich}
  et~al.}{2019}]{Pillepich2019}
{Pillepich} A.,  et~al., 2019, \mn@doi [\mnras] {10.1093/mnras/stz2338}, \href
  {https://ui.adsabs.harvard.edu/abs/2019MNRAS.490.3196P} {490, 3196}

\bibitem[\protect\citeauthoryear{{Planck Collaboration} et~al.,}{{Planck
  Collaboration} et~al.}{2016}]{PlankColaboration2016}
{Planck Collaboration} et~al., 2016, \mn@doi [\aap]
  {10.1051/0004-6361/201525830}, \href
  {https://ui.adsabs.harvard.edu/abs/2016A&A...594A..13P} {594, A13}

\bibitem[\protect\citeauthoryear{{Porciani}, {Dekel}  \& {Hoffman}}{{Porciani}
  et~al.}{2002a}]{Porciani2002a}
{Porciani} C.,  {Dekel} A.,   {Hoffman} Y.,  2002a, \mn@doi [\mnras]
  {10.1046/j.1365-8711.2002.05305.x}, \href
  {https://ui.adsabs.harvard.edu/abs/2002MNRAS.332..325P} {332, 325}

\bibitem[\protect\citeauthoryear{{Porciani}, {Dekel}  \& {Hoffman}}{{Porciani}
  et~al.}{2002b}]{Porciani2002b}
{Porciani} C.,  {Dekel} A.,   {Hoffman} Y.,  2002b, \mn@doi [\mnras]
  {10.1046/j.1365-8711.2002.05306.x}, \href
  {https://ui.adsabs.harvard.edu/abs/2002MNRAS.332..339P} {332, 339}

\bibitem[\protect\citeauthoryear{{Posti}, {Fraternali}  \& {Marasco}}{{Posti}
  et~al.}{2019}]{Posti2019}
{Posti} L.,  {Fraternali} F.,   {Marasco} A.,  2019, \mn@doi [\aap]
  {10.1051/0004-6361/201935553}, \href
  {https://ui.adsabs.harvard.edu/abs/2019A&A...626A..56P} {626, A56}

\bibitem[\protect\citeauthoryear{{Prole}, {van der Burg}, {Hilker}  \&
  {Spitler}}{{Prole} et~al.}{2021}]{Prole2021}
{Prole} D.~J.,  {van der Burg} R.~F.~J.,  {Hilker} M.,   {Spitler} L.~R.,
  2021, \mn@doi [\mnras] {10.1093/mnras/staa3296}, \href
  {https://ui.adsabs.harvard.edu/abs/2021MNRAS.500.2049P} {500, 2049}

\bibitem[\protect\citeauthoryear{{Read}, {Iorio}, {Agertz}  \&
  {Fraternali}}{{Read} et~al.}{2016}]{Read2016b}
{Read} J.~I.,  {Iorio} G.,  {Agertz} O.,   {Fraternali} F.,  2016, \mn@doi
  [\mnras] {10.1093/mnras/stw1876}, \href
  {https://ui.adsabs.harvard.edu/abs/2016MNRAS.462.3628R} {462, 3628}

\bibitem[\protect\citeauthoryear{{Read}, {Iorio}, {Agertz}  \&
  {Fraternali}}{{Read} et~al.}{2017}]{Read2017}
{Read} J.~I.,  {Iorio} G.,  {Agertz} O.,   {Fraternali} F.,  2017, \mn@doi
  [\mnras] {10.1093/mnras/stx147}, \href
  {https://ui.adsabs.harvard.edu/abs/2017MNRAS.467.2019R} {467, 2019}

\bibitem[\protect\citeauthoryear{{Rodriguez-Gomez} et~al.,}{{Rodriguez-Gomez}
  et~al.}{2015}]{RodriguezGomez2015}
{Rodriguez-Gomez} V.,  et~al., 2015, \mn@doi [\mnras] {10.1093/mnras/stv264},
  \href {https://ui.adsabs.harvard.edu/abs/2015MNRAS.449...49R} {449, 49}

\bibitem[\protect\citeauthoryear{{Rom{\'a}n} \& {Trujillo}}{{Rom{\'a}n} \&
  {Trujillo}}{2017a}]{Roman2017a}
{Rom{\'a}n} J.,  {Trujillo} I.,  2017a, \mn@doi [\mnras]
  {10.1093/mnras/stx438}, \href
  {https://ui.adsabs.harvard.edu/abs/2017MNRAS.468..703R} {468, 703}

\bibitem[\protect\citeauthoryear{{Rom{\'a}n} \& {Trujillo}}{{Rom{\'a}n} \&
  {Trujillo}}{2017b}]{Roman2017b}
{Rom{\'a}n} J.,  {Trujillo} I.,  2017b, \mn@doi [\mnras]
  {10.1093/mnras/stx694}, \href
  {https://ui.adsabs.harvard.edu/abs/2017MNRAS.468.4039R} {468, 4039}

\bibitem[\protect\citeauthoryear{{Rom{\'a}n}, {Beasley}, {Ruiz-Lara}  \&
  {Valls-Gabaud}}{{Rom{\'a}n} et~al.}{2019}]{Roman2019}
{Rom{\'a}n} J.,  {Beasley} M.~A.,  {Ruiz-Lara} T.,   {Valls-Gabaud} D.,  2019,
  \mn@doi [\mnras] {10.1093/mnras/stz835}, \href
  {https://ui.adsabs.harvard.edu/abs/2019MNRAS.486..823R} {486, 823}

\bibitem[\protect\citeauthoryear{{Rong}, {Guo}, {Gao}, {Liao}, {Xie}, {Puzia},
  {Sun}  \& {Pan}}{{Rong} et~al.}{2017}]{Rong2017}
{Rong} Y.,  {Guo} Q.,  {Gao} L.,  {Liao} S.,  {Xie} L.,  {Puzia} T.~H.,  {Sun}
  S.,   {Pan} J.,  2017, \mn@doi [\mnras] {10.1093/mnras/stx1440}, \href
  {https://ui.adsabs.harvard.edu/abs/2017MNRAS.470.4231R} {470, 4231}

\bibitem[\protect\citeauthoryear{{Rong}, {Mancera Pi{\~n}a}, {Tempel}, {Puzia}
  \& {De Rijcke}}{{Rong} et~al.}{2020a}]{Rong2020}
{Rong} Y.,  {Mancera Pi{\~n}a} P.~E.,  {Tempel} E.,  {Puzia} T.~H.,   {De
  Rijcke} S.,  2020a, \mn@doi [\mnras] {10.1093/mnrasl/slaa129}, \href
  {https://ui.adsabs.harvard.edu/abs/2020MNRAS.498L..72R} {498, L72}

\bibitem[\protect\citeauthoryear{{Rong}, {Zhu}, {Johnston}, {Zhang}, {Cao},
  {Puzia}  \& {Galaz}}{{Rong} et~al.}{2020b}]{Rong2020b}
{Rong} Y.,  {Zhu} K.,  {Johnston} E.~J.,  {Zhang} H.-X.,  {Cao} T.,  {Puzia}
  T.~H.,   {Galaz} G.,  2020b, \mn@doi [\apjl] {10.3847/2041-8213/aba8aa},
  \href {https://ui.adsabs.harvard.edu/abs/2020ApJ...899L..12R} {899, L12}

\bibitem[\protect\citeauthoryear{{Safarzadeh} \& {Scannapieco}}{{Safarzadeh} \&
  {Scannapieco}}{2017}]{Safarzadeh2017}
{Safarzadeh} M.,  {Scannapieco} E.,  2017, \mn@doi [\apj]
  {10.3847/1538-4357/aa94c8}, \href
  {https://ui.adsabs.harvard.edu/abs/2017ApJ...850...99S} {850, 99}

\bibitem[\protect\citeauthoryear{{Saifollahi}, {Trujillo}, {Beasley},
  {Peletier}  \& {Knapen}}{{Saifollahi} et~al.}{2021}]{Saifollahi2021}
{Saifollahi} T.,  {Trujillo} I.,  {Beasley} M.~A.,  {Peletier} R.~F.,
  {Knapen} J.~H.,  2021, \mn@doi [\mnras] {10.1093/mnras/staa3016}, \href
  {https://ui.adsabs.harvard.edu/abs/2021MNRAS.502.5921S} {502, 5921}

\bibitem[\protect\citeauthoryear{{Saifollahi}, {Zaritsky}, {Trujillo},
  {Peletier}, {Knapen}, {Amorisco}, {Beasley}  \& {Donnerstein}}{{Saifollahi}
  et~al.}{2022}]{Saifollahi2022}
{Saifollahi} T.,  {Zaritsky} D.,  {Trujillo} I.,  {Peletier} R.~F.,  {Knapen}
  J.~H.,  {Amorisco} N.,  {Beasley} M.~A.,   {Donnerstein} R.,  2022, \mn@doi
  [\mnras] {10.1093/mnras/stac328}, \href
  {https://ui.adsabs.harvard.edu/abs/2022MNRAS.511.4633S} {511, 4633}

\bibitem[\protect\citeauthoryear{{Sales}, {Navarro}, {Schaye}, {Dalla Vecchia},
  {Springel}, {Haas}  \& {Helmi}}{{Sales} et~al.}{2009}]{Sales2009}
{Sales} L.~V.,  {Navarro} J.~F.,  {Schaye} J.,  {Dalla Vecchia} C.,  {Springel}
  V.,  {Haas} M.~R.,   {Helmi} A.,  2009, \mn@doi [\mnras]
  {10.1111/j.1745-3933.2009.00725.x}, \href
  {https://ui.adsabs.harvard.edu/abs/2009MNRAS.399L..64S} {399, L64}

\bibitem[\protect\citeauthoryear{{Sales}, {Navarro}, {Schaye}, {Dalla Vecchia},
  {Springel}  \& {Booth}}{{Sales} et~al.}{2010}]{Sales2010}
{Sales} L.~V.,  {Navarro} J.~F.,  {Schaye} J.,  {Dalla Vecchia} C.,  {Springel}
  V.,   {Booth} C.~M.,  2010, \mn@doi [\mnras]
  {10.1111/j.1365-2966.2010.17391.x}, \href
  {https://ui.adsabs.harvard.edu/abs/2010MNRAS.409.1541S} {409, 1541}

\bibitem[\protect\citeauthoryear{{Sales}, {Navarro}, {Theuns}, {Schaye},
  {White}, {Frenk}, {Crain}  \& {Dalla Vecchia}}{{Sales}
  et~al.}{2012}]{Sales2012}
{Sales} L.~V.,  {Navarro} J.~F.,  {Theuns} T.,  {Schaye} J.,  {White} S. D.~M.,
   {Frenk} C.~S.,  {Crain} R.~A.,   {Dalla Vecchia} C.,  2012, \mn@doi [\mnras]
  {10.1111/j.1365-2966.2012.20975.x}, \href
  {https://ui.adsabs.harvard.edu/abs/2012MNRAS.423.1544S} {423, 1544}

\bibitem[\protect\citeauthoryear{{Sales}, {Navarro}, {Pe{\~n}afiel}, {Peng},
  {Lim}  \& {Hernquist}}{{Sales} et~al.}{2020}]{Sales2020}
{Sales} L.~V.,  {Navarro} J.~F.,  {Pe{\~n}afiel} L.,  {Peng} E.~W.,  {Lim} S.,
   {Hernquist} L.,  2020, \mn@doi [\mnras] {10.1093/mnras/staa854}, \href
  {https://ui.adsabs.harvard.edu/abs/2020MNRAS.494.1848S} {494, 1848}

\bibitem[\protect\citeauthoryear{{Sales}, {Wetzel}  \& {Fattahi}}{{Sales}
  et~al.}{2022}]{Sales2022}
{Sales} L.~V.,  {Wetzel} A.,   {Fattahi} A.,  2022, \mn@doi [Nature Astronomy]
  {10.1038/s41550-022-01689-w}, \href
  {https://ui.adsabs.harvard.edu/abs/2022NatAs.tmp..130S} {}

\bibitem[\protect\citeauthoryear{{Sandage} \& {Binggeli}}{{Sandage} \&
  {Binggeli}}{1984}]{Sandage1984}
{Sandage} A.,  {Binggeli} B.,  1984, \mn@doi [\aj] {10.1086/113588}, \href
  {https://ui.adsabs.harvard.edu/abs/1984AJ.....89..919S} {89, 919}

\bibitem[\protect\citeauthoryear{{Scannapieco} et~al.,}{{Scannapieco}
  et~al.}{2012}]{Scannapieco2012}
{Scannapieco} C.,  et~al., 2012, \mn@doi [\mnras]
  {10.1111/j.1365-2966.2012.20993.x}, \href
  {https://ui.adsabs.harvard.edu/abs/2012MNRAS.423.1726S} {423, 1726}

\bibitem[\protect\citeauthoryear{{Schaye} et~al.,}{{Schaye}
  et~al.}{2010}]{Schaye2010}
{Schaye} J.,  et~al., 2010, \mn@doi [\mnras]
  {10.1111/j.1365-2966.2009.16029.x}, \href
  {https://ui.adsabs.harvard.edu/abs/2010MNRAS.402.1536S} {402, 1536}

\bibitem[\protect\citeauthoryear{{Sellwood} \& {Sanders}}{{Sellwood} \&
  {Sanders}}{2022}]{Sellwood2022}
{Sellwood} J.~A.,  {Sanders} R.~H.,  2022, \mn@doi [\mnras]
  {10.1093/mnras/stac1604}, \href
  {https://ui.adsabs.harvard.edu/abs/2022MNRAS.514.4008S} {514, 4008}

\bibitem[\protect\citeauthoryear{{Shi} et~al.,}{{Shi} et~al.}{2017}]{Shi2017}
{Shi} D.~D.,  et~al., 2017, \mn@doi [\apj] {10.3847/1538-4357/aa8327}, \href
  {https://ui.adsabs.harvard.edu/abs/2017ApJ...846...26S} {846, 26}

\bibitem[\protect\citeauthoryear{{Somerville} et~al.,}{{Somerville}
  et~al.}{2018}]{Somerville2018}
{Somerville} R.~S.,  et~al., 2018, \mn@doi [\mnras] {10.1093/mnras/stx2040},
  \href {https://ui.adsabs.harvard.edu/abs/2018MNRAS.473.2714S} {473, 2714}

\bibitem[\protect\citeauthoryear{{Spekkens} \& {Karunakaran}}{{Spekkens} \&
  {Karunakaran}}{2018}]{Spekkens_and_Karunakaran2018}
{Spekkens} K.,  {Karunakaran} A.,  2018, \mn@doi [\apj]
  {10.3847/1538-4357/aa94be}, \href
  {https://ui.adsabs.harvard.edu/abs/2018ApJ...855...28S} {855, 28}

\bibitem[\protect\citeauthoryear{{Springel}}{{Springel}}{2010}]{Springel2010}
{Springel} V.,  2010, \mn@doi [\mnras] {10.1111/j.1365-2966.2009.15715.x},
  \href {https://ui.adsabs.harvard.edu/abs/2010MNRAS.401..791S} {401, 791}

\bibitem[\protect\citeauthoryear{{Springel} \& {Hernquist}}{{Springel} \&
  {Hernquist}}{2003}]{SpringelHernquist2003}
{Springel} V.,  {Hernquist} L.,  2003, \mn@doi [\mnras]
  {10.1046/j.1365-8711.2003.06206.x}, \href
  {https://ui.adsabs.harvard.edu/abs/2003MNRAS.339..289S} {339, 289}

\bibitem[\protect\citeauthoryear{{Springel}, {White}, {Tormen}  \&
  {Kauffmann}}{{Springel} et~al.}{2001}]{Springel2001}
{Springel} V.,  {White} S. D.~M.,  {Tormen} G.,   {Kauffmann} G.,  2001,
  \mn@doi [\mnras] {10.1046/j.1365-8711.2001.04912.x}, \href
  {https://ui.adsabs.harvard.edu/abs/2001MNRAS.328..726S} {328, 726}

\bibitem[\protect\citeauthoryear{{Springel} et~al.,}{{Springel}
  et~al.}{2018}]{Springel2018}
{Springel} V.,  et~al., 2018, \mn@doi [\mnras] {10.1093/mnras/stx3304}, \href
  {https://ui.adsabs.harvard.edu/abs/2018MNRAS.475..676S} {475, 676}

\bibitem[\protect\citeauthoryear{{Toloba} et~al.,}{{Toloba}
  et~al.}{2018}]{Toloba2018}
{Toloba} E.,  et~al., 2018, \mn@doi [\apjl] {10.3847/2041-8213/aab603}, \href
  {https://ui.adsabs.harvard.edu/abs/2018ApJ...856L..31T} {856, L31}

\bibitem[\protect\citeauthoryear{{Tremmel}, {Wright}, {Brooks}, {Munshi},
  {Nagai}  \& {Quinn}}{{Tremmel} et~al.}{2020}]{Tremmel2020}
{Tremmel} M.,  {Wright} A.~C.,  {Brooks} A.~M.,  {Munshi} F.,  {Nagai} D.,
  {Quinn} T.~R.,  2020, \mn@doi [\mnras] {10.1093/mnras/staa2015}, \href
  {https://ui.adsabs.harvard.edu/abs/2020MNRAS.497.2786T} {497, 2786}

\bibitem[\protect\citeauthoryear{{Trujillo}}{{Trujillo}}{2021}]{Trujillo2021}
{Trujillo} I.,  2021, \mn@doi [Nature Astronomy] {10.1038/s41550-021-01388-y},
  \href {https://ui.adsabs.harvard.edu/abs/2021NatAs...5.1182T} {5, 1182}

\bibitem[\protect\citeauthoryear{{Trujillo-Gomez}, {Kruijssen}  \&
  {Reina-Campos}}{{Trujillo-Gomez} et~al.}{2022}]{Trujillo-Gomez2022}
{Trujillo-Gomez} S.,  {Kruijssen} J.~M.~D.,   {Reina-Campos} M.,  2022, \mn@doi
  [\mnras] {10.1093/mnras/stab3401}, \href
  {https://ui.adsabs.harvard.edu/abs/2022MNRAS.510.3356T} {510, 3356}

\bibitem[\protect\citeauthoryear{{{\"U}bler}, {Naab}, {Oser}, {Aumer}, {Sales}
  \& {White}}{{{\"U}bler} et~al.}{2014}]{Ubler2014}
{{\"U}bler} H.,  {Naab} T.,  {Oser} L.,  {Aumer} M.,  {Sales} L.~V.,   {White}
  S. D.~M.,  2014, \mn@doi [\mnras] {10.1093/mnras/stu1275}, \href
  {https://ui.adsabs.harvard.edu/abs/2014MNRAS.443.2092U} {443, 2092}

\bibitem[\protect\citeauthoryear{{Van Nest}, {Munshi}, {Wright}, {Tremmel},
  {Brooks}, {Nagai}  \& {Quinn}}{{Van Nest} et~al.}{2022}]{VanNest2022}
{Van Nest} J.~D.,  {Munshi} F.,  {Wright} A.~C.,  {Tremmel} M.,  {Brooks}
  A.~M.,  {Nagai} D.,   {Quinn} T.,  2022, \mn@doi [\apj]
  {10.3847/1538-4357/ac43b7}, \href
  {https://ui.adsabs.harvard.edu/abs/2022ApJ...926...92V} {926, 92}

\bibitem[\protect\citeauthoryear{{Venhola} et~al.,}{{Venhola}
  et~al.}{2017}]{Venhola2017}
{Venhola} A.,  et~al., 2017, \mn@doi [\aap] {10.1051/0004-6361/201730696},
  \href {https://ui.adsabs.harvard.edu/abs/2017A&A...608A.142V} {608, A142}

\bibitem[\protect\citeauthoryear{{Venhola} et~al.,}{{Venhola}
  et~al.}{2022}]{Venhola2022}
{Venhola} A.,  et~al., 2022, \mn@doi [\aap] {10.1051/0004-6361/202141756},
  \href {https://ui.adsabs.harvard.edu/abs/2022A&A...662A..43V} {662, A43}

\bibitem[\protect\citeauthoryear{{Vogelsberger}, {Genel}, {Sijacki}, {Torrey},
  {Springel}  \& {Hernquist}}{{Vogelsberger} et~al.}{2013}]{Vogelsberger2013}
{Vogelsberger} M.,  {Genel} S.,  {Sijacki} D.,  {Torrey} P.,  {Springel} V.,
  {Hernquist} L.,  2013, \mn@doi [\mnras] {10.1093/mnras/stt1789}, \href
  {https://ui.adsabs.harvard.edu/abs/2013MNRAS.436.3031V} {436, 3031}

\bibitem[\protect\citeauthoryear{{Vogelsberger} et~al.,}{{Vogelsberger}
  et~al.}{2014}]{Vogelsberger2014a}
{Vogelsberger} M.,  et~al., 2014, \mn@doi [Monthly Notices of the Royal
  Astronomical Society] {10.1093/mnras/stu1536}, \href
  {https://ui.adsabs.harvard.edu/abs/2014MNRAS.444.1518V} {444, 1518}

\bibitem[\protect\citeauthoryear{{Wasserman}, {Romanowsky}, {Brodie}, {van
  Dokkum}, {Conroy}, {Abraham}, {Cohen}  \& {Danieli}}{{Wasserman}
  et~al.}{2018}]{Wasserman2018}
{Wasserman} A.,  {Romanowsky} A.~J.,  {Brodie} J.,  {van Dokkum} P.,  {Conroy}
  C.,  {Abraham} R.,  {Cohen} Y.,   {Danieli} S.,  2018, \mn@doi [\apjl]
  {10.3847/2041-8213/aad779}, \href
  {https://ui.adsabs.harvard.edu/abs/2018ApJ...863L..15W} {863, L15}

\bibitem[\protect\citeauthoryear{{Weinberger} et~al.,}{{Weinberger}
  et~al.}{2017}]{Weinberger2017}
{Weinberger} R.,  et~al., 2017, \mn@doi [\mnras] {10.1093/mnras/stw2944}, \href
  {https://ui.adsabs.harvard.edu/abs/2017MNRAS.465.3291W} {465, 3291}

\bibitem[\protect\citeauthoryear{{Weinberger} et~al.,}{{Weinberger}
  et~al.}{2018}]{Weinberger2018}
{Weinberger} R.,  et~al., 2018, \mn@doi [\mnras] {10.1093/mnras/sty1733}, \href
  {https://ui.adsabs.harvard.edu/abs/2018MNRAS.479.4056W} {479, 4056}

\bibitem[\protect\citeauthoryear{{Wheeler} et~al.,}{{Wheeler}
  et~al.}{2017}]{Wheeler2017}
{Wheeler} C.,  et~al., 2017, \mn@doi [\mnras] {10.1093/mnras/stw2583}, \href
  {https://ui.adsabs.harvard.edu/abs/2017MNRAS.465.2420W} {465, 2420}

\bibitem[\protect\citeauthoryear{{White}}{{White}}{1984}]{White1984}
{White} S.~D.~M.,  1984, \mn@doi [\apj] {10.1086/162573}, \href
  {https://ui.adsabs.harvard.edu/abs/1984ApJ...286...38W} {286, 38}

\bibitem[\protect\citeauthoryear{{Wright}, {Tremmel}, {Brooks}, {Munshi},
  {Nagai}, {Sharma}  \& {Quinn}}{{Wright} et~al.}{2021}]{Wright2021}
{Wright} A.~C.,  {Tremmel} M.,  {Brooks} A.~M.,  {Munshi} F.,  {Nagai} D.,
  {Sharma} R.~S.,   {Quinn} T.~R.,  2021, \mn@doi [\mnras]
  {10.1093/mnras/stab081}, \href
  {https://ui.adsabs.harvard.edu/abs/2021MNRAS.502.5370W} {502, 5370}

\bibitem[\protect\citeauthoryear{{Yagi}, {Koda}, {Komiyama}  \&
  {Yamanoi}}{{Yagi} et~al.}{2016}]{Yagi2016}
{Yagi} M.,  {Koda} J.,  {Komiyama} Y.,   {Yamanoi} H.,  2016, \mn@doi [\apjs]
  {10.3847/0067-0049/225/1/11}, \href
  {https://ui.adsabs.harvard.edu/abs/2016ApJS..225...11Y} {225, 11}

\bibitem[\protect\citeauthoryear{{Zaritsky}, {Donnerstein}, {Karunakaran},
  {Barbosa}, {Dey}, {Kadowaki}, {Spekkens}  \& {Zhang}}{{Zaritsky}
  et~al.}{2022}]{Zaritsky2022}
{Zaritsky} D.,  {Donnerstein} R.,  {Karunakaran} A.,  {Barbosa} C.~E.,  {Dey}
  A.,  {Kadowaki} J.,  {Spekkens} K.,   {Zhang} H.,  2022, \mn@doi [\apjs]
  {10.3847/1538-4365/ac6ceb}, \href
  {https://ui.adsabs.harvard.edu/abs/2022ApJS..261...11Z} {261, 11}

\bibitem[\protect\citeauthoryear{{de Blok} \& {McGaugh}}{{de Blok} \&
  {McGaugh}}{1997}]{deBlok1997}
{de Blok} W.~J.~G.,  {McGaugh} S.~S.,  1997, \mn@doi [\mnras]
  {10.1093/mnras/290.3.533}, \href
  {https://ui.adsabs.harvard.edu/abs/1997MNRAS.290..533D} {290, 533}

\bibitem[\protect\citeauthoryear{{van Dokkum}, {Abraham}, {Merritt}, {Zhang},
  {Geha}  \& {Conroy}}{{van Dokkum} et~al.}{2015a}]{vanDokkum2015a}
{van Dokkum} P.~G.,  {Abraham} R.,  {Merritt} A.,  {Zhang} J.,  {Geha} M.,
  {Conroy} C.,  2015a, \mn@doi [\apjl] {10.1088/2041-8205/798/2/L45}, \href
  {https://ui.adsabs.harvard.edu/abs/2015ApJ...798L..45V} {798, L45}

\bibitem[\protect\citeauthoryear{{van Dokkum} et~al.,}{{van Dokkum}
  et~al.}{2015b}]{vanDokkum2015b}
{van Dokkum} P.~G.,  et~al., 2015b, \mn@doi [\apjl]
  {10.1088/2041-8205/804/1/L26}, \href
  {https://ui.adsabs.harvard.edu/abs/2015ApJ...804L..26V} {804, L26}

\bibitem[\protect\citeauthoryear{{van Dokkum} et~al.,}{{van Dokkum}
  et~al.}{2016}]{vanDokkum2016}
{van Dokkum} P.,  et~al., 2016, \mn@doi [\apjl] {10.3847/2041-8205/828/1/L6},
  \href {https://ui.adsabs.harvard.edu/abs/2016ApJ...828L...6V} {828, L6}

\bibitem[\protect\citeauthoryear{{van Dokkum} et~al.,}{{van Dokkum}
  et~al.}{2017}]{vanDokkum2017}
{van Dokkum} P.,  et~al., 2017, \mn@doi [\apjl] {10.3847/2041-8213/aa7ca2},
  \href {https://ui.adsabs.harvard.edu/abs/2017ApJ...844L..11V} {844, L11}

\bibitem[\protect\citeauthoryear{{van Dokkum} et~al.,}{{van Dokkum}
  et~al.}{2018}]{vanDokkum2018}
{van Dokkum} P.,  et~al., 2018, \mn@doi [\nat] {10.1038/nature25767}, \href
  {https://ui.adsabs.harvard.edu/abs/2018Natur.555..629V} {555, 629}

\bibitem[\protect\citeauthoryear{{van Dokkum}, {Danieli}, {Abraham}, {Conroy}
  \& {Romanowsky}}{{van Dokkum} et~al.}{2019a}]{vanDokkum2019a_DF4}
{van Dokkum} P.,  {Danieli} S.,  {Abraham} R.,  {Conroy} C.,   {Romanowsky}
  A.~J.,  2019a, \mn@doi [\apjl] {10.3847/2041-8213/ab0d92}, \href
  {https://ui.adsabs.harvard.edu/abs/2019ApJ...874L...5V} {874, L5}

\bibitem[\protect\citeauthoryear{{van Dokkum} et~al.,}{{van Dokkum}
  et~al.}{2019b}]{vanDokkum2019}
{van Dokkum} P.,  et~al., 2019b, \mn@doi [\apj] {10.3847/1538-4357/ab2914},
  \href {https://ui.adsabs.harvard.edu/abs/2019ApJ...880...91V} {880, 91}

\bibitem[\protect\citeauthoryear{{van den Bosch}, {Ogiya}, {Hahn}  \&
  {Burkert}}{{van den Bosch} et~al.}{2018}]{vandenBosch2018}
{van den Bosch} F.~C.,  {Ogiya} G.,  {Hahn} O.,   {Burkert} A.,  2018, \mn@doi
  [\mnras] {10.1093/mnras/stx2956}, \href
  {https://ui.adsabs.harvard.edu/abs/2018MNRAS.474.3043V} {474, 3043}

\bibitem[\protect\citeauthoryear{{van der Burg}, {Muzzin}  \& {Hoekstra}}{{van
  der Burg} et~al.}{2016}]{vanderBurg2016}
{van der Burg} R. F.~J.,  {Muzzin} A.,   {Hoekstra} H.,  2016, \mn@doi [\aap]
  {10.1051/0004-6361/201628222}, \href
  {https://ui.adsabs.harvard.edu/abs/2016A&A...590A..20V} {590, A20}

\bibitem[\protect\citeauthoryear{{van der Burg} et~al.,}{{van der Burg}
  et~al.}{2017}]{vanderBurg2017}
{van der Burg} R. F.~J.,  et~al., 2017, \mn@doi [\aap]
  {10.1051/0004-6361/201731335}, \href
  {https://ui.adsabs.harvard.edu/abs/2017A&A...607A..79V} {607, A79}

\makeatother
\end{thebibliography}



\appendix

\section{Halo spin and fraction of angular momentum}
\label{app:lambda_and_jd}

\begin{figure}
\centering
\includegraphics[width=\columnwidth]{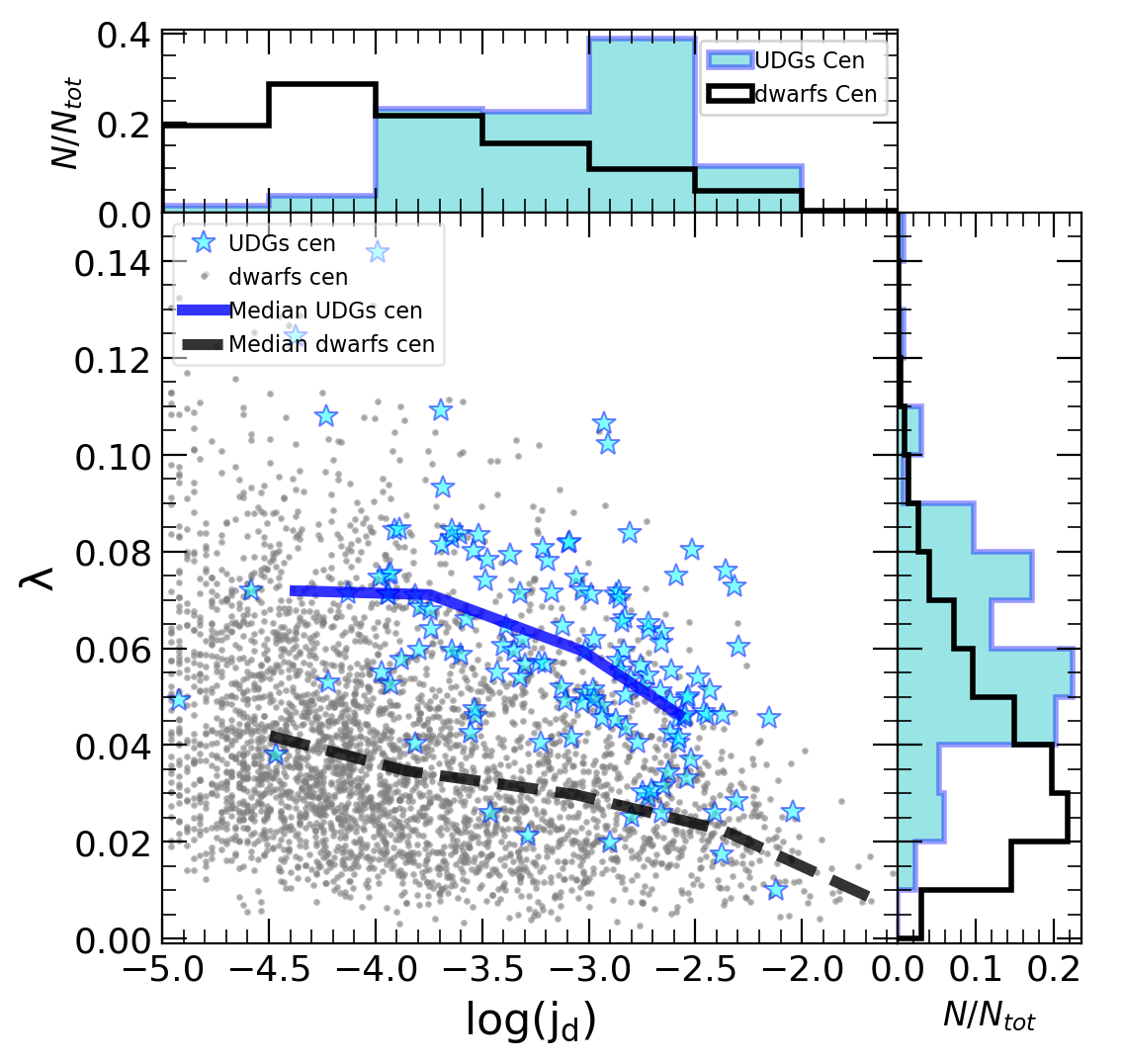}
\caption{Halo spin parameter as a function of the fraction of specific angular momentum. Only central galaxies are shown, grey for normal dwarfs and blue stars for simulated central UDGs (without including the population of backsplash UDGs). Medians for each population are indicated by the dashed black and solid blue lines for normal dwarfs and central UDGs, respectively.}
\label{fig:jd_vs_lambda}
\end{figure}

We find that the sample of simulated UDGs has systematically higher halo spin values than the population of normal dwarf galaxies (see Fig.~\ref{fig:spin}). On the other hand, we show that the central UDGs retain a fraction of the angular momentum of the halo ($j_d$) with respect to normal dwarfs, for a fixed stellar mass fraction ($m_d$) (see Fig.~\ref{fig:mdjd}). Therefore, in Fig.~\ref{fig:jd_vs_lambda} we present the relationship between these two variables, where it can be observed that the central UDGs present systematically higher values at a fixed halo angular momentum fraction.

\section{Mergers in UDGs}
\label{app:mergers}

\begin{figure}
\centering
\includegraphics[width=\columnwidth]{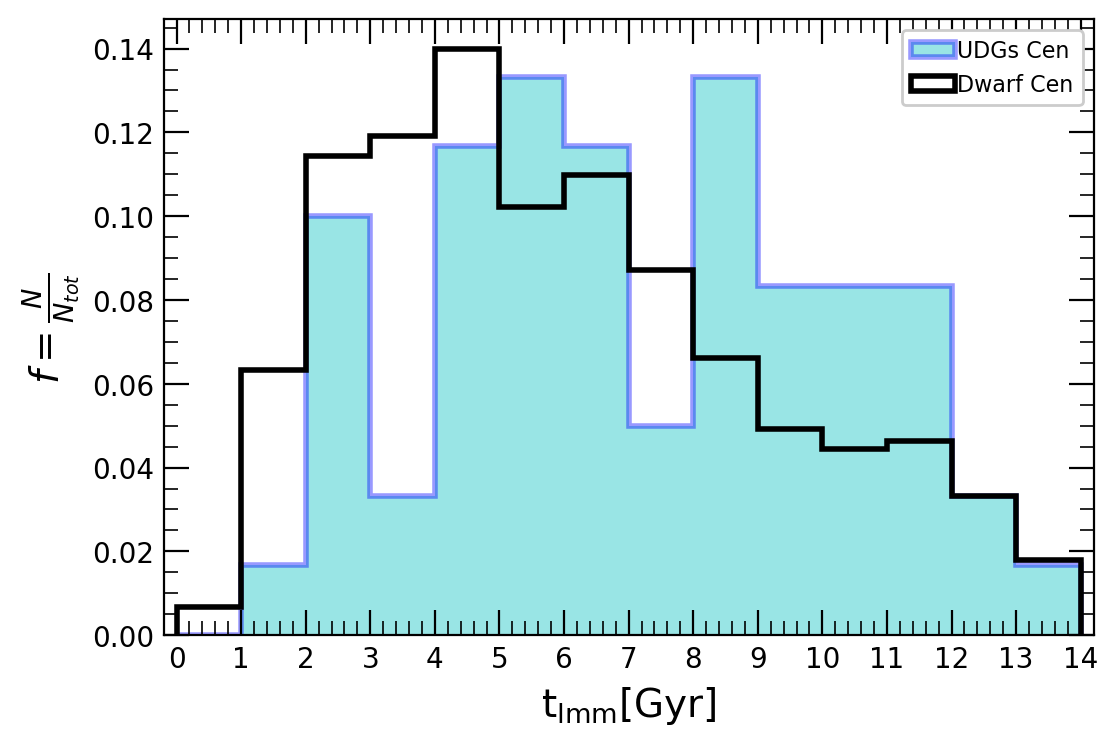}
\caption{Distribution of the time of the last major merger ($\mu_{\star} > 0.2$) for the case of UDGs (blue) and normal dwarfs galaxies (black).}
\label{fig:histo_lmm}
\end{figure}

We find no significant differences in the merger histories of central UDGs and non-UDG central dwarfs. For instance, Fig.~\ref{fig:histo_lmm} shows the time for the last major merger for each sample, where major mergers have been defined as events with stellar mass ratios $\geq 0.2$, following the definition adopted in \citet{Wright2021}. While there is a slight preference for UDGs to have later major merger events than non-UDG dwarfs, we have checked that the time of last major merger does not correspond to individual increases in halo spin or stellar sizes in our UDG sample. Note that the results do not change significantly if a different mass ratio cut is assumed \citep[for instance, $\mu_\star > 0.25$ following ][]{RodriguezGomez2015}.

\bsp	
\label{lastpage}
\end{document}